\documentclass[pdflatex,sn-mathphys-num]{sn-jnl}

\usepackage{graphicx}%
\usepackage{hyperref}
\usepackage{multirow}%
\usepackage{amsmath,amssymb,amsfonts}%
\usepackage{amsthm}%
\usepackage{mathrsfs}%
\usepackage[title]{appendix}%
\usepackage{xcolor}%
\usepackage{textcomp}%
\usepackage{manyfoot}%
\usepackage{booktabs}%
\usepackage{algorithm}%
\usepackage{algorithmicx}%
\usepackage{algpseudocode}%
\usepackage{listings}%
\usepackage{bm}
\usepackage{upgreek}
\usepackage[normalem]{ulem}
\usepackage{esvect}
\usepackage{tabularx}

\newcommand{\call}{{\mathcal{L}}}
\newcommand{\calf}{{\mathcal{F}}}
\newcommand{\fs}{{\mathcal{F}\text{-statistic}}}
\newcommand{\bmB}{{\bm{B}}}
\newcommand{\bmt}{{\bm{\theta}}}
\newcommand{\calz}{{\mathcal{Z}}}
\newcommand{\rmd}{{\rm{d}}}
\newcommand{\apc}{{\varnothing}}
\newcommand{\tpc}{{\otimes}}
\newcommand{\FF}{{FIREFLY}}

\usepackage{geometry}
\theoremstyle{thmstyleone}%
\theoremstyle{thmstyletwo}%

\theoremstyle{thmstylethree}%

\begin{document}

\title[FIREFLY]{A practical Bayesian method for 
gravitational-wave ringdown analysis with multiple modes
}


\author[1,2]{\fnm{Yiming} \sur{Dong}}\email{ydong@pku.edu.cn}
\equalcont{These authors contributed equally to this work.}

\author[1,2]{\fnm{Ziming} \sur{Wang}}\email{zwang@pku.edu.cn}
\equalcont{These authors contributed equally to this work.}

\author[3]{\fnm{Hai-Tian} \sur{Wang}}\email{wanght9@dlut.edu.cn}

\author[4]{\fnm{Junjie} \sur{Zhao}}\email{junjiezhao@hnas.ac.cn}

\author*[2,5]{\fnm{Lijing} \sur{Shao}}\email{lshao@pku.edu.cn}

\affil[1]{\orgdiv{Department of Astronomy, School of Physics}, \orgname{Peking
University}, \orgaddress{\street{No. 5 Yiheyuan Road}, \city{Beijing},
\postcode{100871}, \state{Beijing}, \country{China}}}

\affil[2]{\orgdiv{Kavli Institute for Astronomy and Astrophysics},
\orgname{Peking University}, \orgaddress{\street{No. 5 Yiheyuan Road},
\city{Beijing}, \postcode{100871}, \state{Beijing}, \country{China}}}

\affil[3]{\orgdiv{School of Physics}, \orgname{Dalian University of Technology},
\orgaddress{\street{No. 2 Linggong Road}, \city{Dalian}, \postcode{116024},
\state{Liaoning}, \country{China}}}

\affil[4]{\orgdiv{Institute for Gravitational Wave Astronomy}, \orgname{Henan
Academy of Sciences}, \orgaddress{\street{No. 228 Chongshili Road},
\city{Zhengzhou}, \postcode{450046}, \state{Henan}, \country{China}}}

\affil[5]{\orgdiv{National Astronomical Observatories}, \orgname{Chinese Academy
of Sciences}, \orgaddress{\street{20A Datun Road}, \city{Beijing},
\postcode{100012}, \state{Beijing}, \country{China}}}

\abstract{Gravitational-wave (GW) ringdown signals from black holes (BHs)
encode crucial information about the gravitational dynamics in the strong-field
regime, which offers unique insights into BH properties. In the future, the
improving sensitivity of GW detectors is to enable the extraction of multiple
quasi-normal modes (QNMs) from ringdown signals. However, incorporating
multiple modes drastically enlarges the parameter space, posing computational
challenges to data analysis.  Inspired by the $\mathcal{F}$-statistic method in
the continuous GW searches, we develope an algorithm, dubbed as FIREFLY, for
accelerating the ringdown signal analysis. \FF{} analytically marginalizes the
amplitude and phase parameters of QNMs to reduce the computational cost and
speed up the full-parameter inference from hours to minutes, while achieving
consistent posterior and evidence. The acceleration becomes more significant
when more QNMs are considered.  Rigorously based on the principle of Bayesian
inference and importance sampling, our method is statistically interpretable,
flexible in prior choice, and compatible with various advanced sampling
techniques, providing a new perspective for accelerating future GW data
analysis.}

\maketitle

Since the detection of the first gravitational wave (GW)
event~\cite{LIGOScientific:2016aoc}, we have opened new avenues for advancing
our understanding of gravity~\cite{LIGOScientific:2016lio,
LIGOScientific:2021sio}, cosmology~\cite{LIGOScientific:2017adf}, and
astrophysics of compact stars~\cite{LIGOScientific:2018cki, KAGRA:2021duu}. The
final stage of binary black hole (BH) mergers, known as ringdown, describes the
progression of remnants to the stable phase, offering a unique opportunity to
study BH theories, such as the no-hair theorem~\cite{Isi:2019aib}, the BH area
law~\cite{Isi:2020tac}, and properties of the event
horizon~\cite{Cardoso:2016rao}.  A ringdown signal is usually modelled as the
superposition of quasi-normal modes (QNMs) of the remnant BH, which can be
further divided into spin-weighted spheroidal harmonics with angular indices
$(\ell, m)$. Each set of $(\ell, m)$ contains a series of overtone modes,
indexed by $n$~\cite{Teukolsky:1973ha,Berti:2007zu}. The extraction and
measurement of multiple QNMs provide unparalleled opportunities for studying
BHs,  known as the BH spectroscopy~\cite{Dreyer:2003bv,Berti:2007zu}.

The parameter inference of GW signals is mainly based on the Bayes' theorem, 
\begin{equation}
    P(x| d) = \frac{ P(d |x)P(x)}{P(d)} = \frac{\call(x)\pi(x)}{\calz}\,,
    \label{eq:Bayes theorem}
\end{equation}
where in the second equality, we have introduced the terminologies in the
Bayesian inference: $\pi(x)\equiv P(x)$ is called the \textit{prior},
representing the knowledge of the model parameters $x$ before observing the
data $d$; the \textit{likelihood} $\call(x)\equiv P(d|x)$ is determined by the
model, encoding the statistical relation between the parameter and the data;
the denominator $\calz \equiv P(d)$ is called \textit{evidence}, which can also
be regarded as the normalization factor of the numerator according to the law
of total probability, $\calz = \int \pi(x) \call(x) \rmd x$. The
\textit{posterior} distribution $P(x | d)$ is the conditional probability of
parameters given the data, representing the measurement of parameters after the
observation. Therefore, given the prior and the likelihood, the Bayesian
inference calculates the posterior distribution according to
Eq.~\eqref{eq:Bayes theorem}. However, obtaining an analytical expression (or
an efficient numerical formula) of the posterior is usually unfeasible, and a
more practical approach is to sample from the posterior distribution and
estimate the evidence as an alternative.  There are already some techniques to
carry out this, such as the Markov-Chain Monte Carlo
(MC)~\cite{Christensen:1998gf, Sharma:2017wfu} and nested sampling
methods~\cite{Skilling:2004pqw, Skilling:2006gxv}. These methods partially
reduce the difficulty of posterior calculations, but the computational cost is
still high when sampling in high-dimensional parameter spaces.

Currently, a typical Bayesian analysis of GW signals can take several hours to
days, depending on the number of parameters and the adopted waveform templates.
In the future, the next-generation (XG) ground-based GW detectors, such as the
Cosmic Explorer (CE)~\cite{Reitze:2019iox, Reitze:2019dyk} and the Einstein
Telescope (ET)~\cite{Punturo:2010zz, Hild:2010id}, will be observing with one
order of magnitude higher sensitivity than the current ones, while the
near-future space-based detectors, such as the Laser Interferometer Space
Antenna (LISA)~\cite{LISA:2017pwj}, Taiji~\cite{Hu:2017mde} and
TianQin~\cite{TianQin:2015yph}, are expected to detect GWs in the millihertz
band.  It will become feasible to extract multiple QNMs from ringdown
signals~\cite{Bhagwat:2021kwv, Bhagwat:2023jwv, Pitte:2024zbi}. Each QNM
component is described by a damping harmonic oscillator consisting of four
parameters: amplitude, phase, damping time, and oscillation
frequency~\cite{Berti:2009kk}.  In the General Relativity, the latter two
parameters are determined by the final mass and spin of the remnant BH.
Therefore, adding one QNM mode to ringdown signals increases the dimension of
the parameter space by two, and extracting multiple QNMs leads to a challenge
in the data analysis. The computational costs are even more prohibitive in
tests of no-hair theorem with multiple QNMs, where each mode introduces four
free parameters~\cite{Bhagwat:2021kwv, Bhagwat:2023jwv, Pitte:2024zbi}.

In this work, we propose an algorithm, named FIREFLY ($\mathcal{F}$-statistic
Inspired REsampling For anaLYzing GW ringdown signals), to accelerate the
Bayesian analysis of parameter estimation. This algorithm is inspired by the
$\mathcal{F}$-statistic method in the continuous GW searches, which reduces the
computational cost by maximizing the likelihood over the extrinsic
parameters~\cite{Jaranowski:1998qm}.  In the \FF{}, the amplitude and phase
parameters in the QNMs are analytically marginalized under a specifically
chosen prior, reducing the dimensionality of parameter
space~\cite{Ashok:2024fts,Wang:2024jlz}. Using this auxiliary step, the
inference under the target prior can be efficiently performed by importance
sampling.  In our work, the FIREFLY algorithm reduces the computational time
from several hours to a few minutes, while producing consistent posterior
distribution and evidence as in the traditional full-parameter Bayesian
inference.

\FF{} does not refine the sampling technique itself, instead, it utilizes the
special form of the likelihood and designs more efficient sampling strategies
to accelerate the posterior calculation. Compared to the pure
$\mathcal{F}$-statistic-based methods, which implicitly assume some specific
priors~\cite{Prix:2009tq, Prix2016, Ashok:2024fts, Wang:2024jlz}, \FF{} is
flexible in prior choices because of its efficient importance-sampling design.
Compared to current machine-learning-based accelerating algorithms for Bayesian
GW data analysis~\cite{Gabbard:2019rde, Dax:2021tsq, Dax:2022pxd,
Pacilio:2024qcq}, \FF{} only utilizes the marginalization and resampling
techniques in Bayesian analysis, offering a clear and intuitive statistical
interpretability. Moreover, any future improvements in the stochastic sampling
technique can be incorporated into \FF{}, further accelerating the inference. 

Now we explain the logic of \FF{}. The recorded GW strain of a ringdown signal
is expressed as~\cite{Berti:2009kk,Wang:2024jlz},
\begin{equation}
    h(t) = \sum_{\ell m n} \, \sum_{j = 1,2} B^{\ell m n, j} g_{\ell m n,
    j}(t)\,,\label{eq:reparameterization of QNM signal}
\end{equation}
where $B^{\ell m n, 1} =  A_{\ell mn}\cos\phi_{\ell mn}$ and $B^{\ell m n, 2} =
A_{\ell mn}\sin\phi_{\ell mn}$ are reparameterization of the QNM amplitude
$A_{\ell mn}$ and phase $\phi_{\ell mn}$, while $g_{\ell m n, j}(t)$ depends on
other source parameters, denoted as $\bmt$.  Assuming a stationary and Gaussian
noise, one finds that the likelihood has a Gaussian form with respect to
$\bmB$~\cite{Wang:2024jlz}
\begin{equation}
    \ln \mathcal{L} = {\cal F} - \frac{1}{2} \Big[\big(B^{\mu}-
    \hat{B}^{\mu}\big) M_{\mu\nu} \big(B^{\nu}- \hat{B}^{\nu}\big)  + \langle d
    | d \rangle \Big] \,,
    \label{eq:ll2}
\end{equation}
where $B^\mu$ represents the unified notation for $B^{\ell m n, j}$, $\langle d
| d \rangle$ is noise-weighted  inner product of the observed data $d$, while
$\calf$, $\hat{B}^{\mu}$, and $M_{\mu\nu}$ are functions of $\bmt$ but
independent of $\bmB$ (see Methods for more details).  Based on this
Gaussian-form likelihood, FIREFLY aims to accelerate the Bayesian inference, by
computing the posterior distribution and evidence given the
likelihood~\eqref{eq:ll2} and the prior $\pi(\bmt,\bmB|\tpc)$ in a target
inference problem. 

The \FF{} algorithm mainly composes of two steps: auxiliary inference and
importance sampling.  In the auxiliary inference, the prior
for $\bmt$ is chosen as the same as the target inference, denoted as
$\pi(\bmt)$, while the prior for $\bmB$ is chosen to be independent of $\bmt$
and flat in $\bmB$ with a large-enough range, $\pi(\bmB|\bmt,\apc) = \pi_B$.
Note that we use $\tpc$ and $\apc$ to distinguish the target and auxiliary
priors.  Under the auxiliary prior, the QNM parameters can be analytically
marginalized, leaving an inference problem only involving $\bmt$. The marginal
likelihood for $\bmt$ reads
\begin{equation}
    \begin{aligned}
	\call^m_\varnothing(\bmt)= \pi_B(2\uppi)^N e^{-\frac{1}{2}\langle d
	|d\rangle}\sqrt{{\rm det}\big(M^{-1}\big) }e^{\calf}\,,
    \end{aligned}\label{eq:marginal likelihood}
\end{equation}
where $N$ is the number of QNMs considered in the ringdown waveform.  FIREFLY
firstly draws samples of $\bmt$ with sampling techniques (such as the nested
sampling) based on the marginal likelihood and the prior for $\bmt$. These
samples are denoted by $\{ \bmt_i |\varnothing \}$. The evidence $\calz_\apc$
is also obtained by the adopted sampling technique. The auxiliary inference is
completed by resampling $\bmB$, which only needs to draw one sample $\bmB_i$
for the $i$-th sample $\bmt_i$ from the Gaussian distribution, ${\cal
N}\big(\hat{\bmB}(\bmt_i), M^{-1}(\bmt_i)\big)$; see Methods for details.

In the second step, \FF{} performs importance sampling based on the posterior
samples, $\{\bmt_i, \bmB_i|\apc\}$, and the evidence $\calz_\apc$ produced by
the auxiliary inference. The evidence under the target prior can be estimated
by a MC integration, while for the posterior, a direct importance
sampling weighted by the prior ratio $\pi(\bmt,\bmB|\tpc)/\pi(\bmt,\bmB|\apc)$
may lead to large variance~\cite{Robert2004}, especially when the target and
auxiliary priors are significantly different (see Methods). Therefore, \FF{}
adopts a two-step importance sampling, dealing with $\bmt$ and $\bmB$
separately. First, the sample weights for $\{\bmt_i|\apc\}$ are calculated
according to the marginal-posterior ratio for $\bmt$ between the target and
auxiliary inferences. After the importance resampling for $\bmt$, the sample
$\bmB_i$, assigned to each $\bmt_i$, is drawn from the conditional posterior,
\begin{equation}
    P(\bmB | \bmt, d,\tpc)\propto \pi(\bmB|\bmt,\tpc)e^{-\frac{1}{2}(B^{\mu}-
    \hat{B}^{\mu}) M_{\mu\nu} (B^{\nu}- \hat{B}^{\nu})}\,,
\end{equation}
where the importance-sampling technique is applied once again with a Gaussian
proposer.

\begin{figure}[h]
    \centering
    \includegraphics[width=1\textwidth]{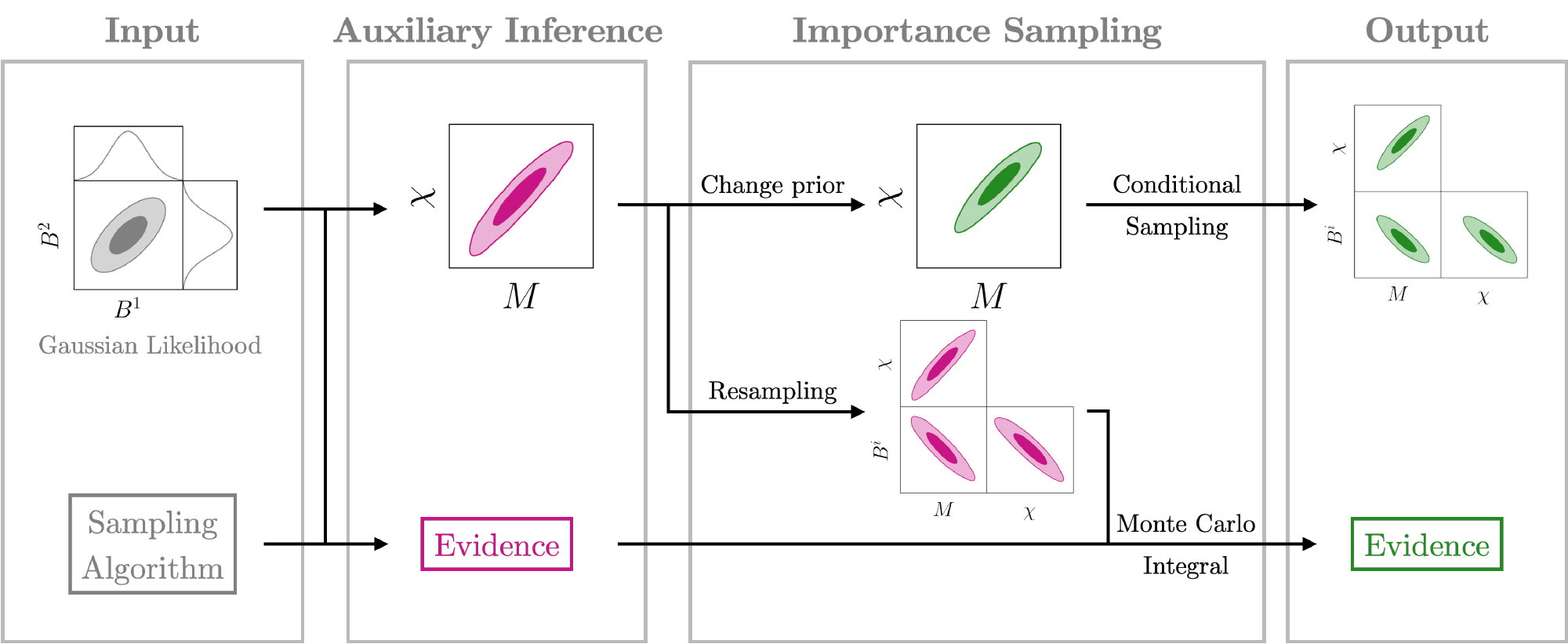}
    \caption{\textbf{Workflow of FIREFLY.} The algorithm takes advantages of
    the special form of the likelihood, where in the ringdown-signal analysis,
    the likelihood is Gaussian with respect to some QNM parameters (represented
    by $\bm{B}$).  Taking this specific likelihood and some state-of-the-art
    sampling techniques as input,  \FF{} consists of two main steps: auxiliary
    inference and importance sampling. In the auxiliary inference, a specific
    prior for $\bm{B}$ is chosen, allowing an analytical marginalization of
    $\bm{B}$. The inference problem involving other parameters (represented by
    $M$ and $\chi$ in the figure) is passed to a stochastic sampling algorithm,
    where in this work we use {\sc dynesty}. The posterior samples and evidence
    under this auxiliary prior are marked in pink. In the importance sampling,
    we use two steps dealing with $\{M,\chi\}$ and $\bm{B}$ respectively. This
    special treatment makes use of the special dependence of likelihood on
    $\bm{B}$, leading to smaller weight variances and more efficient sampling.
    The final output of  \FF{} is the posterior samples and the evidence under
    the target prior, presented in green. }
\label{fig1: workflow}
\end{figure}

The output of \FF{} is the posterior samples $\{\bmt_i,\bmB_i|\tpc\}$ and the
evidence $\calz_\tpc$ under the target prior, which is the same as in the full
Bayesian inference with all parameters. As we can see, \FF{} takes advantage of
the Gaussian-form likelihood in $\bmB$, and eliminates $2N$ parameters
analytically before applying the time-consuming stochastic sampling techniques.
We summarize the workflow of \FF{} in Fig.~\ref{fig1: workflow}.

Now we apply \FF{} to analyze realistic ringdown signals, and compare the
results with those from the full Bayesian inference with all parameters, which
we refer as the ``full-parameter sampling''. To test the performance of \FF{}
in accelerating the extraction of multiple QNMs, we assume the signals are
recorded in the triangular-shape ET which consists of three
detectors~\cite{Hild:2010id}. The data are generated from Numerical Relativity
(NR) waveforms of Simulating eXtreme Spacetimes (SXS)
catalog~\cite{Boyle:2019kee} and injected into a Gaussian stationary noise
corresponding to the ET-D sensitivity. 

Specifically, we inject the NR signal SXS:BBH:0305, which is generated by a
GW150914-like, non-precessing binary BH (BBH) with a mass ratio of $1.2$, a
detector-frame final mass of $68.2\,{M_{\odot}}$, and a dimensionless spin
of $0.69$~\cite{Boyle:2019kee}. The luminosity distance, inclination angle, and
reference phase are set to $390\,{\rm Mpc}$, $3\uppi/4$, and $0$, respectively.
The signal-to-noise ratio of ringdown is approximately $312$ in the ET,
assuming that the signal's starting time is at its peak
amplitude~\cite{Wang:2024jlz}.  Usually, for ringdown signals from nearly
equal-mass BBHs, one focuses solely on the extraction of QNMs with $l=|m|=2$,
while considering different number of overtones~\cite{Giesler:2019uxc}.  In
this work, we consider three scenarios: (i) only the fundamental mode ($N=1$),
(ii) one additional overtone mode ($N=2$), and (iii) two additional overtone
modes ($N=3$).  Since higher-order overtones decay faster and are only
potentially significant at earlier times~\cite{Giesler:2019uxc}, we use the
ringdown part from the whole signal at starting times of $18\,\rm{M}$,
$23\,\rm{M}$ and $28\,\rm{M}$  (with the final mass $\rm{M}=68.2\,{
M_{\odot}}$) after coalescence for $N=1$, $2$ and $3$,
respectively.\footnote{Here we adopt the geometric units $G=c=1$, and
$68.2\,{M}_{\odot}$ corresponds to $0.33\,{\rm ms}$.} In each scenario, it was verified that
reasonable adjustments of the starting time do not significantly change our
results~\cite{Wang:2024jlz}.

When extracting QNMs from the ringdown signal, the variable parameters in the
waveform include the right ascension $\alpha$, declination $\delta$,
polarization angle $\psi$, inclination angle $\iota$, coalescence time $t_c$,
final mass $M_f$, and final dimensionless spin $\chi_f$, in addition to the QNM
parameters $\bmB$.  It is also customary to fix the first five parameters based
on a more comprehensive study, such as the inspiral-merger-ringdown analysis.
For simplicity, we adopt the sky-averaged antenna pattern functions over
$\alpha$, $\delta$, and $\psi$, and fix $\iota$ and $t_c$ to their injected
values. Therefore, in this work $\bmt$ only contains the final mass $M_f$ and
final spin $\chi_f$ of the remnant BH, and the full set of variables is
$\big\{M_f, \chi_f, A_{\ell mn}, \phi_{\ell mn}\big\}$, where $(\ell, m, n)$
runs over all  QNMs considered in the ringdown signal. Notably, expanding the
set of free parameters does not affect the applicability of FIREFLY. In the
target inference, we choose flat priors for all parameters, $M_f/M_{\odot}\sim
{\cal U}(50,100)$, $\chi_f\sim {\cal U}(0,0.99)$, $\phi_{\ell mn}\sim {\cal
U}(0,2\uppi)$ and $A_{\ell mn}\sim {\cal U}(0,A_{\max})$ with $A_{\max}=5\times
10^{-20}$. In the auxiliary inference, the prior for $M_f$ and $\chi_f$ remains
the same, while the flat prior in $\bmB$ corresponds to $\pi\big(A_{\ell mn},
\phi_{\ell mn}\big|\bmt,\apc\big) \propto A_{\ell mn}$. We adopt this prior
shape for all $A_{\ell mn}$ and $\phi_{\ell mn}$, and normalize it in the
region of $A_{\ell mn}\leq A_{\max}$ and $0\leq \phi_{\ell mn}<2\uppi$. 

\begin{figure}[h]
    \centering
    \includegraphics[width=0.95\textwidth]{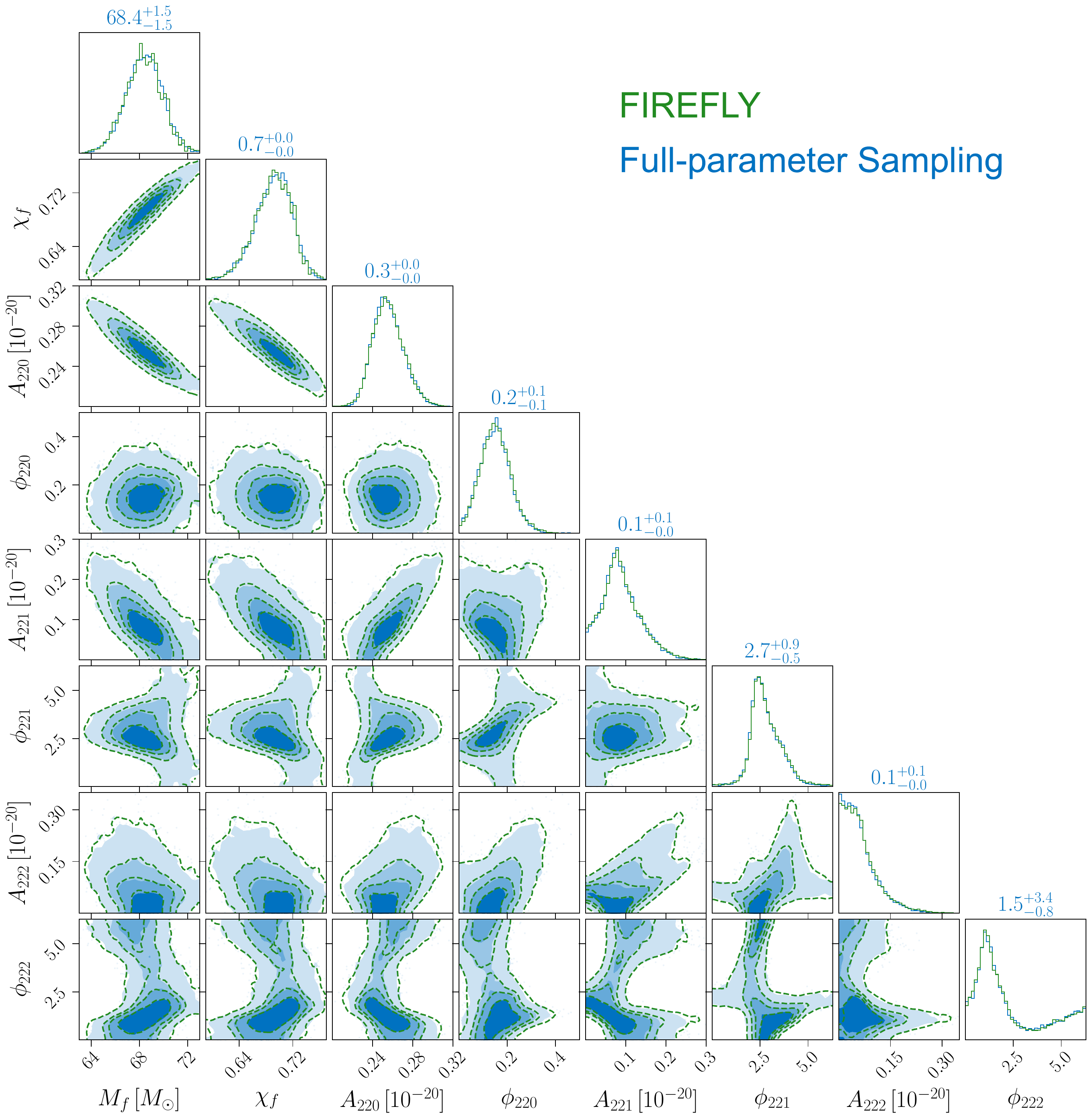}
    \caption{\textbf{Posteriors of the $N=3$ scenario in overtone analysis.}
    The green dashed contours represent the posterior distributions obtained by
    FIREFLY, while the blue ones are from the full-parameter sampling. The
    contours are drawn at the two-dimensional 1-$\sigma$ ($39.3\%$),
    1.5-$\sigma$ ($67.5\%$), 2-$\sigma$ ($86.5\%$), and 3-$\sigma$ ($98.9\%$)
    credible levels.}
\label{fig2: posterior N=3}
\end{figure}
\begin{figure}[h]
    \centering
    \includegraphics[width=1\textwidth]{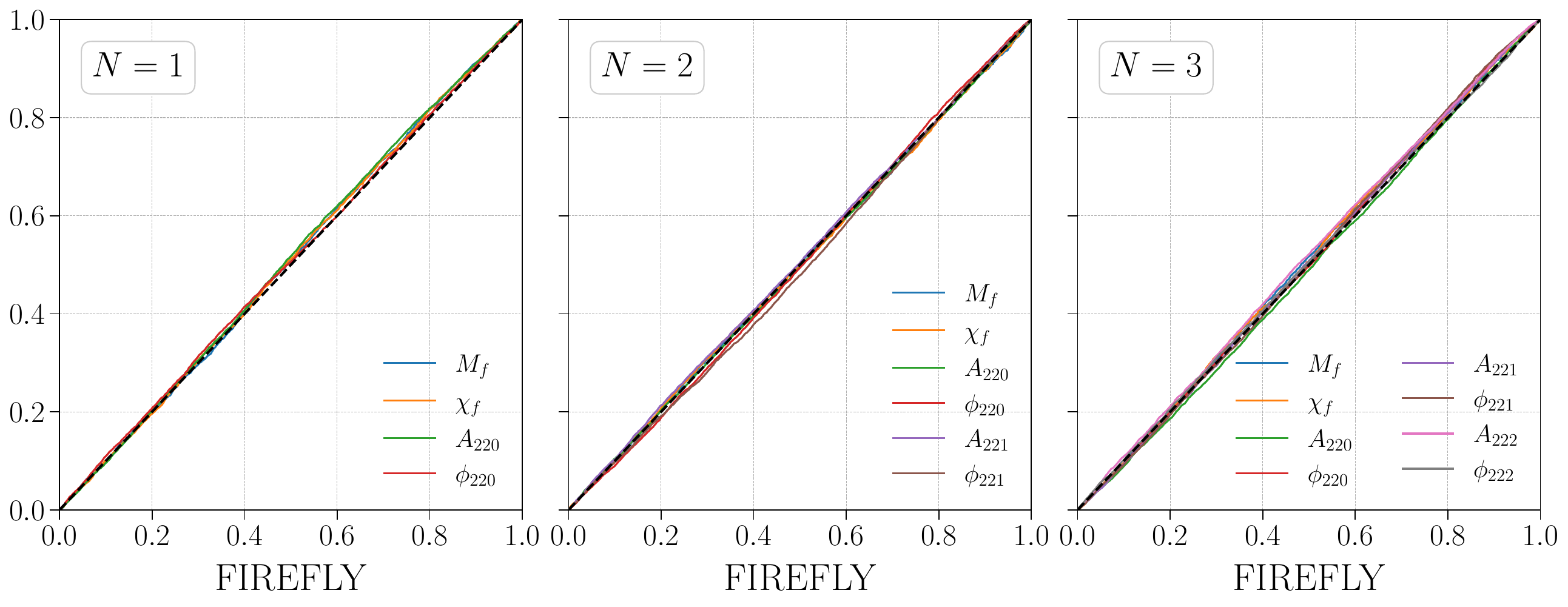}
    \caption{\textbf{P-P plots in overtone analysis.} The differences of the
    one-dimensional marginalized posteriors between FIREFLY and the
    full-parameter sampling are shown in P-P plots. From left to right, the
    panels correspond to $N=1$, $N=2$, and $N=3$ scenarios, respectively.
    Different colors indicate different parameters, while the black dashed
    lines depict $y=x$ for reference.}
\label{fig3: P-P plot}
\end{figure}

Now we compare the posteriors obtained by \FF{}  and the full-parameter
sampling, which directly samples all parameters $\{\bmt, \bmB\}$ under the
target prior. In both the auxiliary inference and the full-parameter sampling,
we use the {\sc dynesty}~\cite{Higson2019} sampler to draw samples from the
posterior.  In all three scenarios, we find that the \FF{} results closely
agree with those from the full-parameter sampling.  For concreteness, we show
the $N=3$ case in Fig.~\ref{fig2: posterior N=3}, where the green and blue
contours represent the results from the FIREFLY and the full-parameter
sampling, respectively. The P-P plots between the one-dimensional marginalized
posteriors of the two methods are shown in Fig.~\ref{fig3: P-P plot}. We also
use the Wasserstein distance to quantitatively verify the consistency
between the two methods~\cite{MAL-073}, and find that the average difference of one-dimensional marginal distributions is less than 0.1 standard deviation.
 The
computational times of \FF{} and the full-parameter sampling are listed in
Table~\ref{tab1: runtime}. In the $N=3$ case,  FIREFLY reduces the
computational time from several hours to just a few minutes, speeding up by two
orders of magnitude. 

\begin{table}[h]
    \renewcommand{\arraystretch}{1.2}
    \caption{\textbf{Computational times with FIREFLY and the full-parameter
    sampling.} The runtime of  FIREFLY consists of three part: ``Sampling''
    denotes the time spent on the nested sampling in auxiliary inferences,
    ``Evidence'' refers to the time to compute the evidence using MC
    integration, and ``IS'' corresponds to the time spent on the importance
    sampling. ``Total'' represents the overall time taken for the entire
    analysis. In the ``Full-Parameter'' column, ``Total'' indicates the time
    spent on the nested sampling over the full parameter space. The $k$-factor
    gives the ratio of computational times used by the full-parameter sampling
    to FIREFLY.}
    \begin{tabular*}{\textwidth}{@{\extracolsep\fill}lllllll}
    \toprule%
    & \multicolumn{4}{@{}c@{}}{FIREFLY} & \multicolumn{1}{@{}c@{}}{Full-parameter} & \multicolumn{1}{@{}c@{}}{} \\\cmidrule(lr){2-5}\cmidrule(lr){6-6}%
     & Sampling & Evidence & IS & Total & Total & $k$-factor \\
    \midrule
    Overtone $N=1$  & ${\rm 45.7\,s}$ & ${\rm 4.5\,s}$ & ${\rm 4.5\,s}$ & ${\rm 54.6\,s}$ & ${\rm 2\,min\ 52.7\,s}$ & 3.2 \\
    Overtone $N=2$  & ${\rm 1\,min\ 7.0\,s}$ & ${\rm 4.8\,s}$ & ${\rm 5.9\,s}$ & ${\rm 1\,min\ 17.7\,s}$ & ${\rm 27\,min\ 59.0\,s}$ & 21.6 \\
    Overtone $N=3$  & ${\rm 2\,min\ 49.7\,s}$ & ${\rm 5.7\,s}$ & ${\rm 7.9\,s}$ & ${\rm 3\,min\ 3.2\,s}$ & ${\rm 5\,hr\ 15\,min\ 38.0\,s}$ & 103.4 \\
    \midrule
    Higher modes  & ${\rm 4\,min\ 30.1\,s}$ & ${\rm 7.2\,s}$ & ${\rm 10.0\,s}$ & ${\rm 4\,min\ 47.2\,s}$ & ${\rm 5\,hr\ 14\,min\ 31.2\,s}$ & 65.7 \\
    \botrule
    \end{tabular*}
    \label{tab1: runtime}
\end{table}

The main reason why \FF{} significantly accelerates the inference is the
analytical marginalization of the QNM amplitude and phase parameters in the
auxiliary inference, which eliminates $2N$ free parameters in the stochastic
sampling process. The computational advantage of  \FF{} becomes more pronounced
when involving more QNMs in the analysis. Due to the same reason,  \FF{} is
fully statistically interpretable, and also compatible with various sampling
techniques. Since the majority of the computational time in \FF{} is spent on
the auxiliary inference (see Table~\ref{tab1: runtime}), one can incorporate
the state-of-the-art sampling techniques into  \FF{}, further reducing the
computational cost.

FIREFLY also allows flexible prior choices, which is based on the effective
two-step importance sampling. In the ringdown signal analysis, different prior
choices can significantly affect posteriors, especially when considering
multiple QNMs or analyzing weak signals (see Methods).  In contrast, the pure
$\mathcal{F}$-statistic-based methods reduce the number of parameters by
analytical maximization, which inherently assumes some specific
priors~\cite{Prix:2009tq, Prix2016, Ashok:2024fts, Wang:2024jlz}. The
flexibility of priors in \FF{} makes it possible to conduct a more
comprehensive analysis when extracting multiple QNMs from ringdown signals. It
is also worth noting that changing to a new target prior only requires one to
rerun an importance sampling that takes only a few seconds, while the results
of auxiliary inference can be reused.



Because of its structure, FIREFLY also gives the evidence estimates, and we
compare them with those from the full-parameter sampling in Fig.~\ref{fig4: log
evidence}.  FIREFLY estimates the evidence with a smaller variance than in the
full-parameter sampling. This is because that the uncertainties of the evidence
estimation mainly come from the stochastic sampling of the auxiliary inference
in \FF, which has $2N$ fewer free parameters than in the full-parameter
sampling.  Nevertheless, the evidence estimated by \FF{} tends to be slightly
larger than that of the full-parameter sampling.  Similar discrepancies have
been reported by other approaches designed to accelerate evidence calculation
in GW Bayesian inference~\cite{Dax:2022pxd}. To address this issue, we conduct
a comprehensive investigation into the potential uncertainty source and confirm
the consistency of the results produced by two methods after considering the
additional uncertainties from the insufficient sampling in the full-parameter
sampling (see Appendix). Having said that, the relative differences between the
two methods are at the level of $10^{-3}$, so the evidence given by \FF{} is
still largely consistent with the full-parameter sampling.

\begin{figure}[h]
    \centering
    \includegraphics[width=0.5\textwidth]{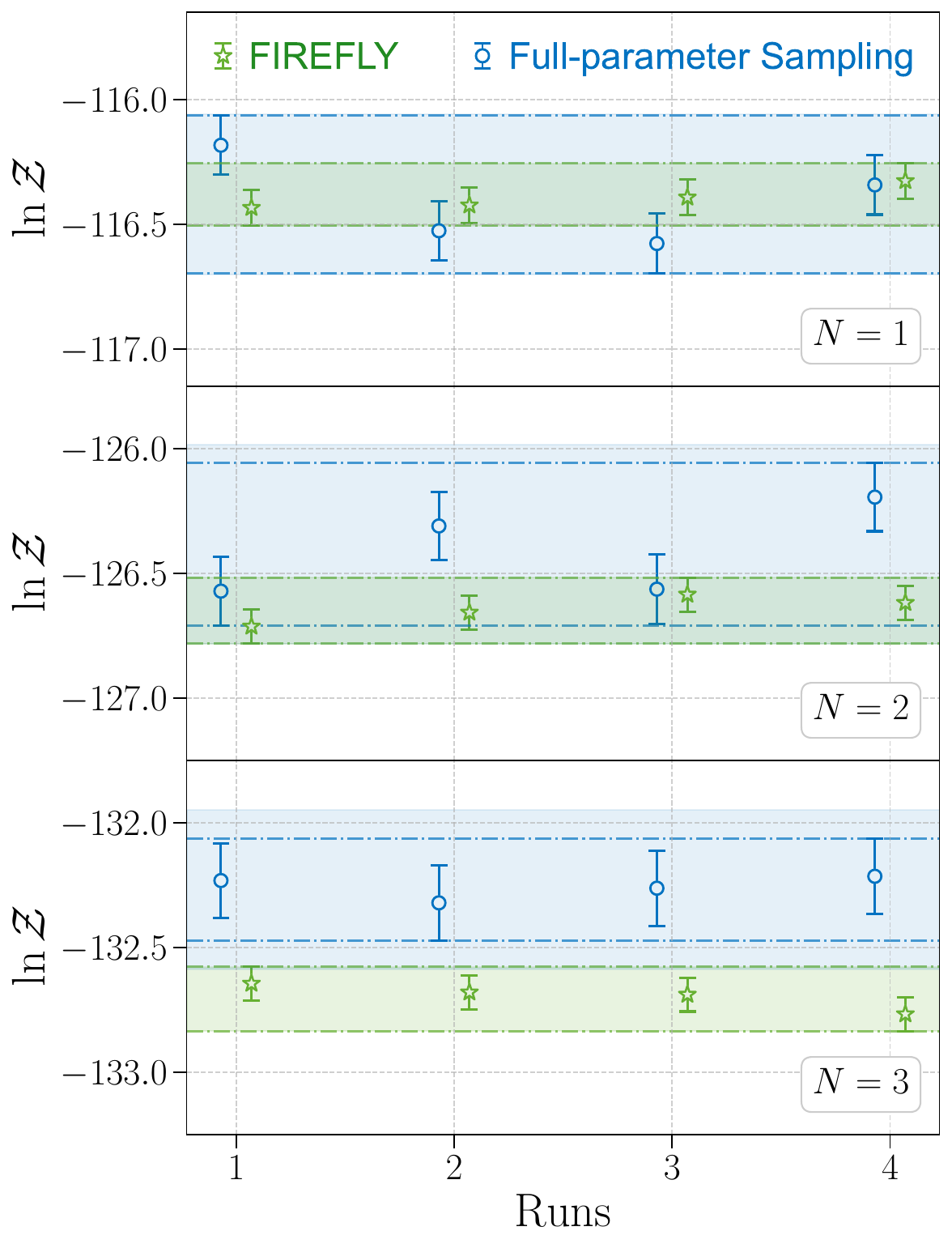}
    \caption{\textbf{Evidences given by FIREFLY and the full-parameter
    sampling.} From top to bottom, we show the evidences for the cases of
    $N=1$, $N=2$, and $N=3$ in overtone analysis, respectively. We use green to
    represent the FIREFLY results and blue for the full-parameter sampling. In
    each case, we independently run four trials, and mark the estimates and
    uncertainties in the figure.  For each method, the dash-dotted lines denote
    the upper and lower bounds of the combined evidence ranges from the four
    trials, while the shaded areas represent conservative evidence ranges
    considering the additional uncertainties arising from insufficient
    sampling.}
\label{fig4: log evidence}
\end{figure}


After validating \FF{} in the parameter estimation of ringdown signals, we now
present the extraction of higher modes in the ringdown signal from an
asymmetric-mass-ratio BBH system. We inject the NR signal SXS:BBH:0065, which
is generated from a non-precessing binary with a mass ratio of $8.0$, a final
redshifted mass of $122.45\,{ M_{\odot}}$, and a final spin of
$0.66$~\cite{Boyle:2019kee}. The luminosity distance, inclination angle, and
reference phase are set to $390\,{\rm Mpc}$, $3\uppi/4$, and $0$, respectively.
We extract four QNMs $(\ell, m, n) = (2,2,1)$, $(2,1,1)$, $(3,3,1)$, and
$(4,4,1)$ from the ringdown signal, at different start times $\Delta t =
16\,{\rm M}$, $20\,{\rm M}$, $24\,{\rm M}$, and $28\,{\rm M}$ with ${\rm
M}=122.45\,{ M_{\odot}}$.\footnote{In the geometric units, ${\rm
M}=122.45\,{ M_{\odot}}$ corresponds to $0.60\,{\rm ms}$.} The prior for the
final mass is chosen to be $M_{f}/M_{\odot}\sim {\cal U}(100,150)$, while
priors of other parameters are the same as those in the previous overtone
analysis.  The joint posteriors of the final mass and the final spin from \FF{}
and the full-parameter sampling are shown in Fig.~\ref{fig5: M-chi higher
modes}.  FIREFLY achieves close agreement with the full-parameter sampling,
while reducing the computational time by about sixty times (see the last row in
Table~\ref{tab1: runtime}).

\begin{figure}[h]
    \centering
    \includegraphics[width=0.5\textwidth]{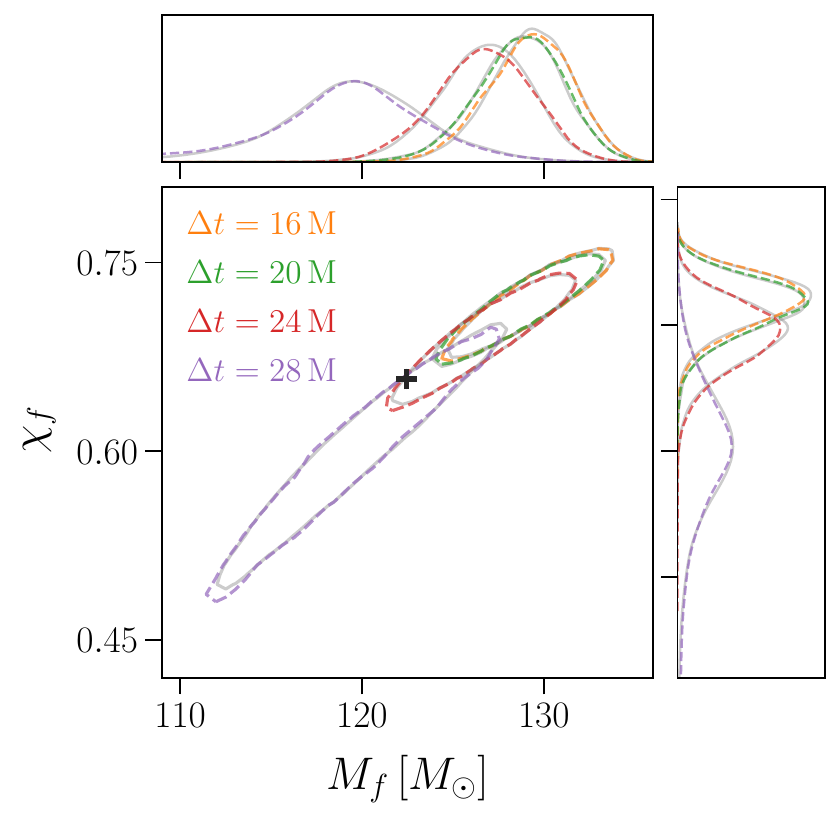}
    \caption{\textbf{Joint Posteriors of the final mass and the final spin with
    different starting times in the higher-mode analysis.} The colored contours
    represent the FIREFLY results, while different colors correspond to
    different starting times. The gray contours indicate results from the
    full-parameter sampling for comparison. The contours are drawn at the
    two-dimensional 2-$\sigma$ ($86.5\%$) credible level.}
\label{fig5: M-chi higher modes}
\end{figure}


In this work, we develop the \FF{} algorithm, which significantly accelerates
the analysis of ringdown signals while maintaining nearly indistinguishable
posterior and evidence estimates. In the spirit of ${\cal F}$-statistic, by
constructing an auxiliary inference where the QNM amplitudes and phases are
analytically marginalized,  \FF{} reduces the computational time by several
orders of magnitude, demonstrating its potential to address the challenge of
dimensionality curse when extracting multiple QNMs. Also, based on the
delicately designed importance-sampling strategy,  \FF{} is highly flexible in
prior choices, allowing for a more comprehensive analysis for identifying
overtones and higher modes in the ringdown signals. \FF{} presents a
computational paradigm that only leverages the intrinsic structure of the
likelihood to accelerate the Bayesian inference, which is fully statistically
interpretable, and compatible with any improvements in stochastic sampling
techniques. Similar structures are commonly encountered in GW analyses, such as
for extreme-mass-ratio inspirals (EMRIs)~\cite{Wang:2012xh} and GW
localization~\cite{Hu:2023hos}. We hope that the novelty of the FIREFLY
algorithm will provide a viable solution to address many more data challenges
in future GW data analysis and the beyond.




\clearpage

\section*{Methods}

\noindent\textbf{Ringdown waveform, likelihood and the $\fs$.}
The ringdown signal of a BBH coalescence models the GW waveform from a
perturbed BH remnant, and is decomposed into the superposition of
QNMs~\cite{Berti:2009kk},
\begin{equation}
    \begin{aligned}
	h_+(t) = \sum_{\ell m n} \! {}_{-2}Y_{\ell m}(\iota, \varsigma) A_{\ell
	mn}e^{-\frac{t}{\tau_{\ell mn}}}\cos \big(2\uppi f_{\ell mn} t +
	\phi_{\ell mn} \big),\\
	h_\times (t) = \sum_{\ell m n} \! {}_{-2}Y_{\ell m}(\iota, \varsigma)
	A_{\ell mn}e^{-\frac{t}{\tau_{\ell mn}}}\sin \big(2\uppi f_{\ell mn} t +
	\phi_{\ell mn} \big),
    \end{aligned}
\label{eq: ringdown QNM}
\end{equation}
where the damping frequency $f_{\ell mn}$ and damping time $\tau_{\ell mn}$ are
determined by the final mass $M_f$ and the final spin $\chi_f$ of the remnant,
$A_{\ell mn}$ and $\phi_{\ell mn}$ are the amplitude and phase of the mode
$(\ell, m,n)$. The binary inclination angle is denoted by $\iota$, while
$\varsigma$ is the azimuthal angle and is usually set to zero. We take the
spin-weighted spherical harmonics ${}_{-2}Y_{\ell m}$ to approximate the
spin-weighted spheroidal harmonics, include the mirror modes, and omit the mode
mixing contributions~\cite{Berti:2014fga, Giesler:2019uxc}.  Combining the
waveform model with the antenna pattern function $F^{+,\times}(\alpha, \delta,
\psi)$, the recorded GW strain can be expressed as the linear summation in
Eq.~\eqref{eq:reparameterization of QNM signal}, where the explicit form of
$g_{\ell m n, j}(t)$ is given by
\begin{equation}
    \begin{aligned}
	g_{\ell m n, 1}(t;\bmt) =&  \Big[F^+\cos(2\uppi f_{\ell mn} t) +
	F^\times\sin(2\uppi f_{\ell mn} t)\Big] {}_{-2}Y_{\ell m}(\iota, 0)
	e^{-\frac{t}{\tau_{\ell mn}}} \,,\\
	g_{\ell m n, 2}(t;\bmt) =&  \Big[ F^\times\cos(2\uppi f_{\ell mn}
	t)-F^+\sin(2\uppi f_{\ell mn} t)\Big] {}_{-2}Y_{\ell m}(\iota, 0)
	e^{-\frac{t}{\tau_{\ell mn}}}\,,
    \end{aligned}
\end{equation}
where $\bmt = \big\{\alpha, \delta, \psi, \iota, M_f, \chi_f \big\}$ comprises
other parameters in the ringdown waveform except for the QNM amplitude and
phase.

In the data analysis of ringdown signals, the likelihood is the conditional
probability of the observed data $d$ given the model $h$, or equivalently the
model parameters $\bmB$ and $\bmt$. In this work we use the standard likelihood
in the GW data analysis, where the log-likelihood reads~\cite{Finn:1992wt}
\begin{equation}
    \ln \call=-\frac{1}{2}\langle d - h | d - h \rangle +C_0\,,\label{eq:GW
    orignal likelihood}
\end{equation}
where $C_0$ does not affect the inference, and is fixed to zero. The inner
product between two signals, $\langle \cdot | \cdot \rangle$, is defined by, 
\begin{equation}
    \langle {h}_1\vert
    {h}_2\rangle=\vv{h}_1^{\intercal}\mathcal{C}^{-1}\vv{h}_2\,,
\end{equation}
where the arrow notation $\vv{h}$ indicates that it is a discrete time series
sampled from the continuous signal $h(t)$, and $\mathcal{C}$ is the
auto-covariance matrix of the noise at the sampling times, which is determined
by the auto-correlation function of the detector noise, or equivalently, the
power spectral density of noise. Equation~\eqref{eq:ll2} is obtained by
formulating the likelihood (\ref{eq:GW orignal likelihood}) into a quadratic
form with respect to $\bmB$, where $M_{\mu\nu} \! = \langle g_\mu | g_\nu
\rangle$ and $\hat{B}^{\mu}\! = (M^{-1})^{\mu\nu} s_{\nu}$ with $s_{\nu} \! =
\langle d | g_\nu \rangle$. The first term in Eq.~\eqref{eq:ll2} is called the
$\fs$,
\begin{equation}
    {\cal F}= \frac{1}{2} s_{\mu} (M^{-1})^{\mu\nu} s_{\nu}\,.\label{eq:fs}
\end{equation}
In addition, if a $N_{\rm det}$-detector network is considered, $s_\mu$ and
$M_{\mu\nu}$ should be replaced by $\sum_{k=1}^{N_{\rm det}} s_{\mu}^{k}$ and
$\sum_{k=1}^{N_{\rm det}} M_{\mu\nu}^{k}$ respectively, where $s_{\mu}^{k} =
\langle d^{k} | g_\mu^k \rangle$ and $M_{\mu\nu}^{k} = \langle g_\mu^k |
g_\nu^k \rangle$ are the corresponding quantities in the $k$-th detector.  As a
short summary, the reparameterization of QNM parameters leads to a
Gaussian-form likelihood in $\bmB$, which is the foundation of our FIREFLY
algorithm.


\noindent\textbf{Auxiliary inference and the marginal likelihood.} In the
auxiliary inference, the posterior distribution is 
\begin{equation}
    P(\bmt,\bmB | d,\apc) = \frac{\pi(\bmt,\bmB|\apc)
    \call(\bmt,\bmB)}{\calz_\apc} = \frac{\pi(\bmt) \pi_B
    \call(\bmt,\bmB)}{\calz_\apc}\,,
\end{equation}
where $\calz_\apc$ is the evidence under the auxiliary prior. Formally, one can
always marginalize over $\bmB$, and rewrite the marginal posterior of $\bmt$ as
\begin{equation}
    \begin{aligned}
      P(\bmt | d,\apc) =
      \frac{\pi(\bmt|\apc)\call^m_\varnothing(\bmt)}{\calz_\apc}\,, \quad\quad
      \calz_\apc =\! \int \pi(\bmt|\apc)\call^m_\varnothing(\bmt) \rmd \bmt\,,
    \end{aligned}\label{eq:posterior of theta with marginal likelihood}
 \end{equation}
 with the marginal likelihood~\cite{Loredo:2024bayesian} 
 \begin{equation}
    \call^m_\varnothing(\bmt) := \int \pi(\bmB|\bmt,\apc) \call(\bmt,\bmB) \rmd
    \bmB\,.
 \end{equation}
 In  \FF{}, the marginal likelihood $\call^m_\varnothing(\bmt)$ is analytically
 given by Eq.~\eqref{eq:marginal likelihood}, since the likelihood
 $\call(\bmt,\bmB)$ is quadratic in $\bmB$ and  $\pi(\bmB|\bmt,\apc)$ is flat
 in $\bmB$ in our auxiliary prior. Benefiting from this, the marginal
 likelihood $\call^m_\varnothing(\bmt)$ can be evaluated at almost the same
 speed as for the full likelihood. Noting that Eq.~\eqref{eq:posterior of theta
 with marginal likelihood} represents an inference problem only involving
 $\bmt$, \FF{} applies current stochastic sampling techniques to draw samples
 based on the marginal likelihood $\call^m_\varnothing(\bmt)$ and the prior
 $\pi(\bmt)$, and the posterior samples can be regarded as drawn from the
 marginal posterior $P(\bmt | d,\apc)$, while producing the same evidence
 $\calz_\apc$ as in the full auxiliary inference. To complete the auxiliary
 inference, we also need to sample $\bmB$, whose conditional posterior given
 $\bmt$ is Gaussian,
 \begin{equation}
     \begin{aligned}
	 P(\bmB | \bmt, d,\apc)
	 = \frac{\pi(\bmB|\bmt,\varnothing)
	 \call(\bmt,\bmB)}{\call^m_\varnothing(\bmt)}=\frac{\exp\left
	 \{-\frac{1}{2}\big(B^{\mu}- \hat{B}^{\mu}\big) M_{\mu\nu}
	 \big(B^{\nu}- \hat{B}^{\nu}\big)\right \}}{(2\uppi)^{N}\sqrt{{\rm
	 det}\big(M^{-1}\big)}}\,.
     \end{aligned}\label{eq:conditional posterior of B}
 \end{equation}
Therefore, for each sample $\bmt_i$, the corresponding $\bmB_i$ is generated by
drawing once from the Gaussian distribution ${\cal N}\big(\hat{\bmB}(\bmt_i),
M^{-1}(\bmt_i)\big)$. A more detailed description of this resampling technique
can be found in the Appendix C of Ref.~\cite{Thrane:2018qnx}.


\noindent\textbf{Evidence estimation and the two-step importance sampling.}
Based on the $n_\apc$ posterior samples $\{ \bmt_i,\bmB_i|\apc \}$ and the
evidence $\calz_\apc$ in the auxiliary inference, \FF{} calculates the evidence
in the target inference $\calz_\tpc$ by the following formula,
\begin{equation}
    \begin{aligned}
	\calz_\tpc &= \int \pi(\bmt,\bmB|\tpc) \call(\bmt,\bmB) \rmd \bmt \rmd
	\bmB 
	=\calz_\apc\int\frac{\pi(\bmB|\tpc)}{\pi(\bmB|\apc)}P(\bmt,\bmB |
	d,\apc) \rmd \bmt \rmd \bmB
	\approx \frac{\calz_\apc}{n_\apc}\sum_{\big\{\bmt_i,\bmB_i|\apc \big\}}
	\frac{\pi(\bmB_i|\tpc)}{\pi(\bmB_i|\apc)}\,.
    \end{aligned}
\label{eq: evidence}
\end{equation} 
In the second equality, the Bayes' theorem is applied to rewrite the evidence
into an integral involving the posterior and evidence in the auxiliary
inference. In the last equality, we approximate the integral by a MC summation
with the posterior samples $\big\{\bmt_i,\bmB_i|\apc\big\}$. This technique is
widely used to calculate the evidence for hyperparameters in the hierarchical
Bayesian inference~\cite{Thrane:2018qnx}.

As for the posterior samples, \FF{} first reweighs $\big\{\bmt_i|\apc\big\}$
according to the marginal posterior density ratio of $\bmt$ between the target
and auxiliary priors, called the marginal importance weight $w^m(\bmt)$,
\begin{equation}
    w^m(\bmt) = \frac{P(\bmt|d,\tpc)}{P(\bmt|d,\apc)} =
    \frac{\pi(\bmt)}{\pi(\bmt)}
    \frac{\call^m_\tpc(\bmt)}{\call^m_\apc(\bmt)}\frac{\calz_\apc}{\calz_\tpc}\,,
\end{equation}
where $\call^m_\tpc(\bmt) = \int \pi(\bmB|\bmt,\tpc) \call(\bmt,\bmB) \rmd
\bmB$ is the marginal likelihood for $\bmt$ under the target prior.
Substituting the marginal likelihood in Eq.~\eqref{eq:marginal likelihood}, we
find that
\begin{equation}
  w^m(\bmt)  \propto \frac{\int \pi(\bmB|\bmt,\tpc) e^{-\frac{1}{2}(B^{\mu}-
  \hat{B}^{\mu}) M_{\mu\nu} (B^{\nu}- \hat{B}^{\nu})} \rmd \bmB}{\sqrt{{\rm
  det}\big(M^{-1}\big) }}\,.
\label{eq: IS weight}
\end{equation}
The integral in the numerator can be efficiently evaluated by a MC integration
with the Gaussian proposer ${\cal N}\big(\hat{\bmB}(\bmt_i),
M^{-1}(\bmt_i)\big)$.  \FF{} resamples $\bmt$ from the weighted sample set
$\big\{ \bmt_i, w^m(\bmt_i)\big\}$ and obtains $n_\tpc$ resampled samples
$\big\{\bmt_i|\tpc\big\}$, which are regarded as being drawn from the marginal
posterior $P(\bmt|d,\tpc)$ according to the principle of importance sampling.
In the second step, samples of $\bmB$ are drawn from the conditional posterior
$P(\bmB|d,\bmt,\tpc)$, which can also be performed by the importance-sampling
technique. For each $\bmt_i \in \{\bmt_i|\tpc\}$,  \FF{} firstly draws $n_{\rm
B}$ samples $\{\bmB_{j|i}\}$ from ${\cal N}\big(\hat{\bmB}(\bmt_i),
M^{-1}(\bmt_i)\big)$, then calculates the weight $w_{j|i} =
\pi(\bmB_{j|i}|\bmt_i,\tpc)$. The resampled $\bmB_i$ assigned to $\bmt_i$ is
generated by a single draw from the weighted sample set $\big\{\bmB_{j|i},
w_{j|i} \big\}$. After repeating this procedure for all $\bmt_i \in
\{\bmt_i|\tpc\}$, $n_\tpc$ samples, denoted as
$\big\{\bmt_i,\bmB_i|\tpc\big\}$, are obtained, which can be regarded as being
drawn from the posterior distribution $P(\bmt,\bmB|d,\tpc)$. 

The strategy of two-step importance sampling allows  \FF{} to efficiently
resample the posterior samples under the target prior. In principle, one can
directly calculate the marginal likelihood $\call^m_\tpc(\bmt)$ by a MC
integration, eliminate the dependence on $\bmB$, and sample the posterior of
$\bmt$ in the target prior. However, this is equivalent to numerically
marginalizing $\bmB$ in the posterior distribution that requires a MC
integration for every likelihood evaluation and is thus computationally
expensive.  In contrast, FIREFLY executes the MC integration \textit{after} the
sampling and on the samples $\{\bmt_i|\apc\}$, making it much more efficient.
In the second step, the proposal distribution for $\bmB$ is Gaussian, which is
also highly efficient for sampling with current algorithms. The two-step
strategy again utilizes the Gaussian form of the likelihood in $\bmB$, reducing
the computational cost and avoiding the weight divergence in the importance
sampling.


\noindent\textbf{The problem of weight divergence.} 
Importance sampling estimates the properties of a target distribution $p(x)$ by
reweighing samples drawn from a proposal distribution $q(x)$~\cite{Tokdar2010}.
After drawing samples $\{x_i\}$ from $q(x)$, in order to get the samples that
follow the target distribution $p(x)$, one computes the importance weights,
\begin{equation}
    w_i = \frac{p(x_i)}{q(x_i)}\,,
\end{equation}
and reweighs the samples $x_i$ accordingly. A handy proposal distribution
should resemble the target distribution, ensuring importance weight close to
one for each sample. If the chosen proposal distribution differs significantly
from the target one, it is likely that some samples will have large importance
weights $\gg 1$, while some others will have weights $\ll 1$, leading to a
large divergence.  The weight divergence refers to a large variance in the
importance weights of samples~\cite{Robert2004}.  In this case, the performance
of importance sampling deteriorates. Intuitively, a few samples with extremely
large weights dominate the importance sampling process, causing significant
instability in the MC estimation.

In  \FF{} we need to convert the posterior of the auxiliary inference to the
posterior with the target prior, where the importance-sampling proposal and
target are the posterior distributions under the auxiliary and target priors.
Since they share the same likelihood, the importance weights are proportional
to the ratio of the priors. When the target prior is significantly different
from the auxiliary one (flat in $\bmB$ in \FF{}), the importance weights can be
highly divergent. In this work, the importance weight takes the form of $w
\propto A_{\ell mn}^{-1}$. Therefore, the samples with small $A_{\ell mn}$ will
have disproportionately large importance weights, leading to severe weight
divergence. A direct importance sampling according to the prior ratio
$\pi(\bmt,\bmB|\tpc)/\pi(\bmt,\bmB|\apc)$ produces unstable distributions with
significant variance and spurious structures, as shown in Fig.~\ref{figM1:
naiveIS}.  \FF{} addresses this issue by performing a two-step importance
sampling, avoiding the direct resampling for all parameters.  In the first
step, \FF{} resamples $\bmt$ with the marginal likelihood ratio $w^m(\bmt)$,
where the problem of weight divergence does not exist. In the second step, one
needs to sample $\bmB$ in the conditional posterior given $\bmt$, which is the
product of the conditional prior $\pi(\bmB|\bmt,\tpc)$ and the Gaussian
distribution ${\cal N}\big(\hat{\bmB}(\bmt), M^{-1}(\bmt)\big)$. Benefiting
from this specific form of the conditional posterior, one can adopt the
importance-sampling technique again to draw samples with a Gaussian proposer.
Though  weight divergence may still arise with some choices of
$\pi(\bmB|\bmt,\tpc)$, the efficient sampling from the Gaussian distribution
effectively mitigates this issue. Overall, the two-step importance sampling
produces a more stable posterior with smaller variances, which is also closer
to the results of full-parameter sampling (cf. Fig.~\ref{fig2: posterior N=3}).

\begin{figure}[h]
    \centering
    \includegraphics[width=1\textwidth]{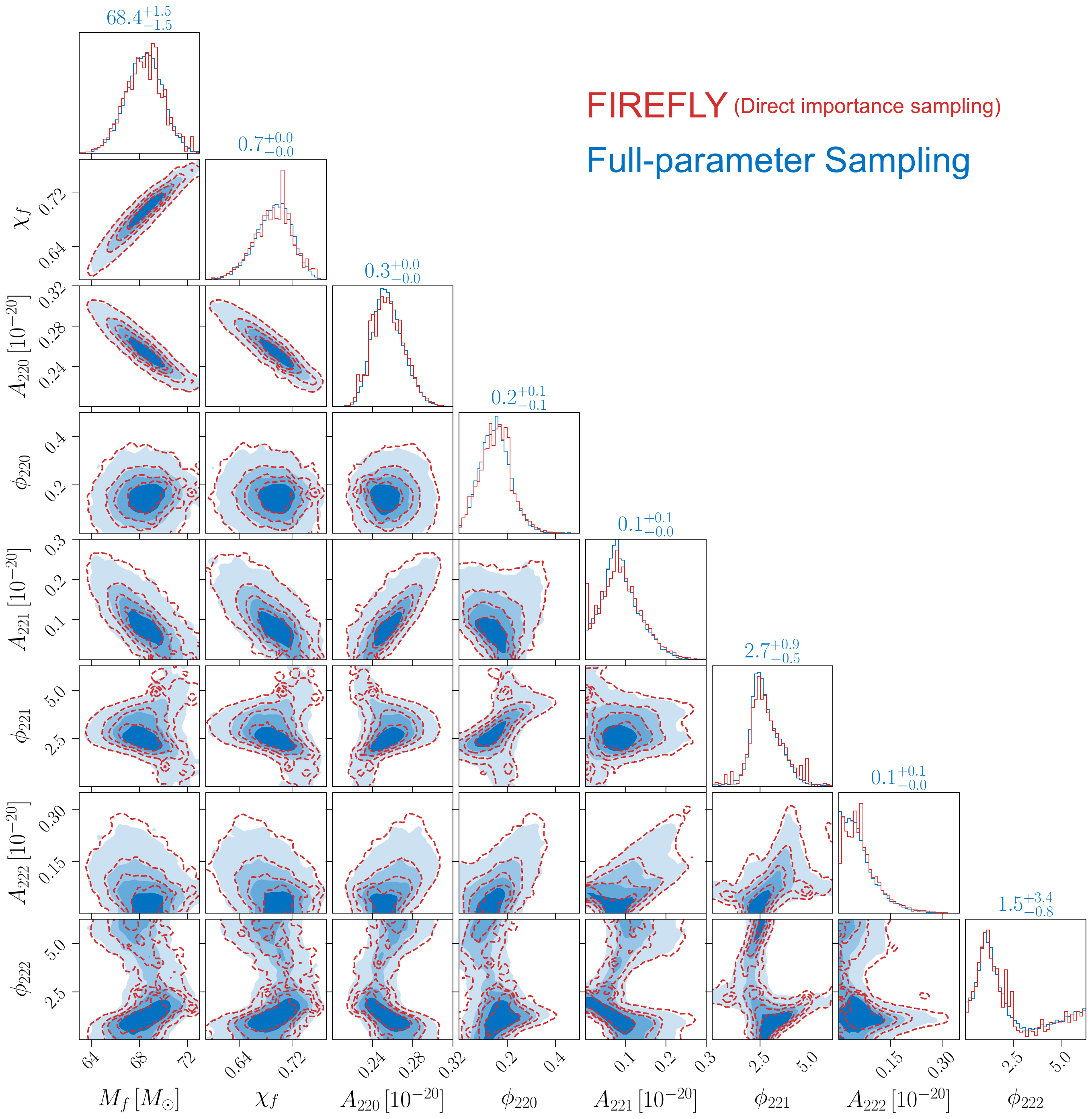}
    \caption{\textbf{Posterior of the $N=3$ scenario from the direct importance
    sampling.} Same as Fig.~\ref{fig2: posterior N=3}, but now the red contours
    represent the posterior distributions obtained through direct importance
    sampling for all parameters.}    
\label{figM1: naiveIS}
\end{figure}

\noindent\textbf{Simulation configuration and hyperparameters.}
We perform nested sampling with the \textsc{Bilby}
package~\cite{Ashton:2018jfp} and the \textsc{dynesty}
sampler~\cite{Higson2019}. The sampler is configured with ${\rm nlive =2000}$
live points and a termination criterion of ${\rm dlogz=0.1}$. In all samplings,
we utilize 16 CPU cores for computation on the FusionServer X6000 V6 equipped
with two Intel Xeon Platinum 8358 processors and 256\,GB of RAM. The sampling
rate of detectors is set to 2048\,Hz, which is sufficient for the ringdown
signals from BH remnants with mass around $100M_\odot$.

Apart from the configurations in the auxiliary inference, \FF{} introduces
several additional hyperparameters. The first hyperparameter $n_{\rm MC}$
defines the number of repetitions for computing the evidence with the MC
integration. In this work, we set $n_{\rm MC}=10$, and calculate the mean and
standard deviation of the evidence over 10 repetitions. In the overtone
analysis, the contribution to the final uncertainty of the evidence in the MC
integration is significantly smaller than that in the auxiliary inference,
where the evidence is estimated by the nested sampling.

The second hyperparameter $n_{w}$ defines the number of samples used to
calculate the marginal importance weights with MC integration in the first
importance sampling step. Our tests show that $n_{w}=5\times 10^{4}$ provides
sufficient accuracy, while increasing $n_{w}$ further does not lead to
significant improvements in the results. We adopt this setting throughout the
work.

The third hyperparameter $n_\tpc$ is the number of samples generated in the
first importance-sampling step, which is also the total number of posterior
samples in  \FF{}. We draw $n_\tpc=2\times 10^{4}$ samples based on the
importance sampling weights in Eq.~(\ref{eq: IS weight}). Subsequently, the
second importance sampling step is then performed to recover the corresponding
parameters $\bmB$ for each sample. 

The fourth hyperparameter $n_{\rm QNM}=5000$ describes the number of samples
drawn from ${\cal N}\big(\hat{\bmB}(\bmt_i), M^{-1}(\bmt_i)\big)$ in the second
importance-sampling step. In our configuration, the number of samples obtained
from the auxiliary prior is smaller than $n_\tpc=2\times 10^{4}$. Therefore,
the effective number of samples for the parameters $\bmt$ is fewer than
$n_\tpc$.  However, when recovering the parameters $\bmB_i$ for each sample
$\bmt_i$, we fully utilize the special form of the conditional posterior by
drawing non-repetitive parameters $\bmB_i$ for each sample $\bmt_i$. Increasing
$n_\tpc$ helps acquire more stable posterior distributions for the parameters
$\bmB$.

By construction of the \FF{} package, the hyperparameters can be adjusted to
either speed up the inference or improve the accuracy and stability of the
results. Additionally, if auxiliary inference becomes sufficiently efficient,
we could explore different importance sampling strategies to generate more
compact posterior sample sets. Under the current configuration, FIREFLY already
delivers sufficiently fast, stable, and accurate results, but further
improvements can be pursued.

\begin{figure}[h]
    \centering
    \includegraphics[width=1\textwidth]{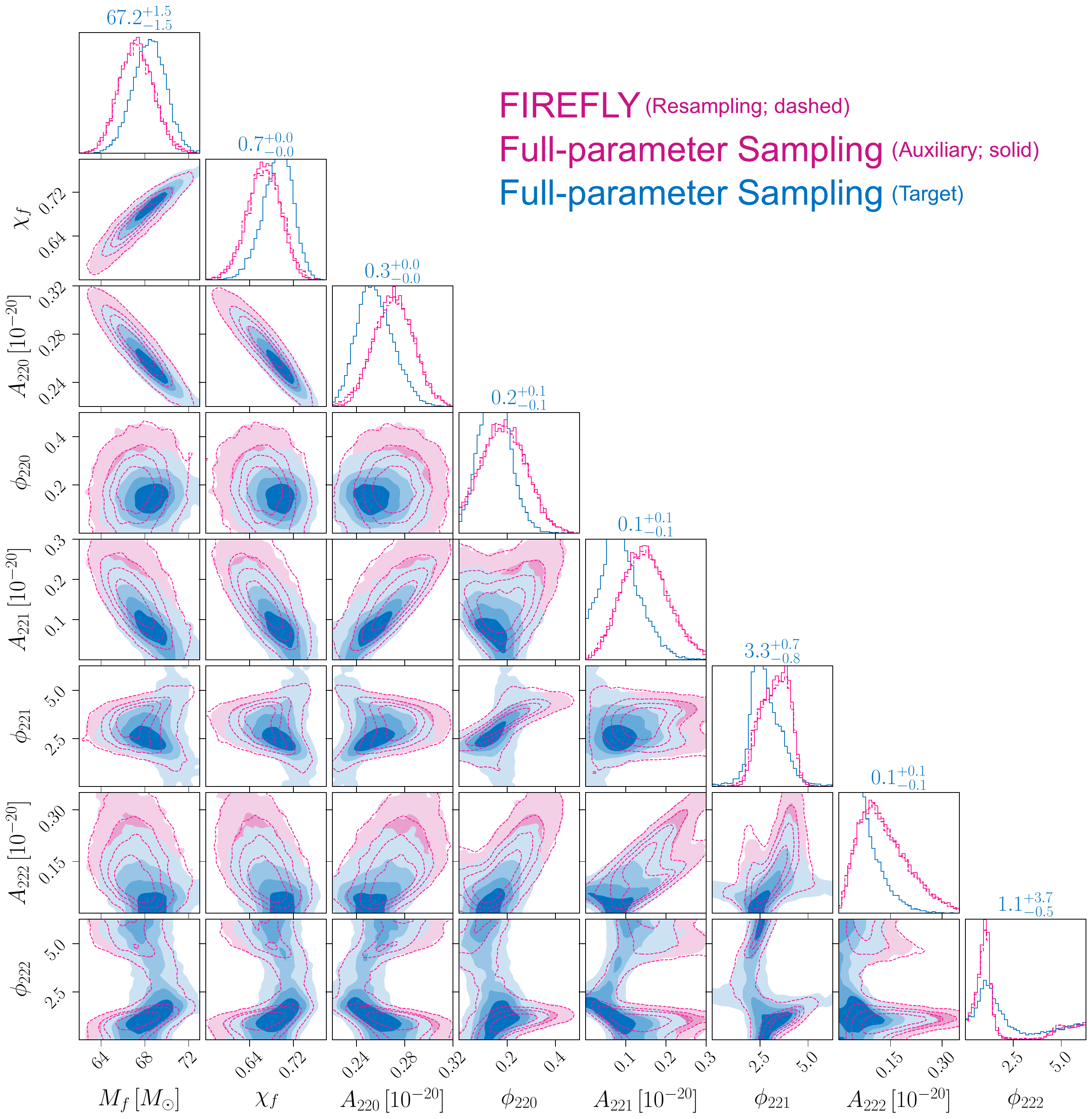}
    \caption{\textbf{Posterior of the $N=3$ scenario under auxiliary prior.}
    Same as Fig.~\ref{fig2: posterior N=3}, but the pink contours represent the
    posteriors under the auxiliary prior. The dashed pink lines are obtained by
    resampling the QNM parameters $\bmB$ based on the posterior samples of
    $\bmt$ in the auxiliary inference of FIREFLY, denoted as ``\FF{}
    (Resampling)'', while the solid pink contours are obtained by the
    full-parameter sampling under the auxiliary prior. The blue contours still
    represent the posterior distributions under the target prior, given by
    FIREFLY.}
\label{figM2: Underlying Prior 223}
\end{figure}

\noindent\textbf{Posterior differences between the auxiliary and target priors.}
\FF{} analytically marginalizes the QNM parameters $\bmB$ in the auxiliary
inference, where the prior $\pi_B$ is chosen to be flat. The full posterior can
be obtained by resampling the QNM parameters $\bmB$ based on the posterior
samples of $\bmt$. In Fig.~\ref{figM2: Underlying Prior 223}, we illustrate the
result of full-parameter sampling under this auxiliary prior for comparison.
For conciseness, we only show the $N=3$ case in overtone analysis. As expected,
the posterior distributions from the resampling of $\bmB$ in \FF{} and the
full-parameter sampling under the auxiliary prior highly agree with each other.
We also illustrate how different priors for $\bmB$ affect the posteriors in the
ringdown analysis, where noticeable differences are observed in
Fig.~\ref{figM2: Underlying Prior 223}.  Compared to the target prior, the
posterior of the amplitudes under the auxiliary prior is overestimated, which
comes from the flat prior in $\bmB$ favoring larger amplitudes. Additionally,
since there is a negative correlation between amplitudes and the final
mass/spin, the overestimation of amplitudes leads to an underestimation of the
mass/spin parameter. For the higher-mode analysis where four QNMs are involved,
the posterior differences between the target and auxiliary priors are more
significant. Therefore, in the ringdown analysis, different prior choices lead
to different posterior distributions, especially extracting multiple QNMs.
Advantageously, in \FF{} the flexibility of the prior selection is achieved by
adopting the two-step importance sampling.

\backmatter

\bmhead{Acknowledgements}
We thank Hyung Won Lee, Hiroyuki Nakano, and Nami Uchikata for comments.

\bmhead{Funding}
This work was supported by the National Natural Science Foundation of China
(123B2043), the Beijing Natural Science Foundation (1242018), the National SKA
Program of China (2020SKA0120300), the Max Planck Partner Group Program funded
by the Max Planck Society, and the High-performance Computing Platform of Peking
University. H.-T.\ Wang and L.\ Shao are supported by ``the Fundamental
Research Funds for the Central Universities'' respectively at Dalian University
of Technology and Peking University. J.\ Zhao is supported by the National Natural
Science Foundation of China (No. 12405052) and the Startup Research Fund of
Henan Academy of Sciences (No. 241841220).

\bmhead{Competing interests}
The authors declare no competing interests.

\bmhead{Data availability}

The data used in this study can be found in the public SXS Gravitational
Waveform Database at \url{https://data.black-holes.org/}.  The data generated
and analysed during this study are available from the corresponding author upon
reasonable request.

\bmhead{Code availability}

The code can be addressed to the corresponding author upon reasonable requests,
and will be made publicly available soon.


\bmhead{Author contributions}
Y.D.\ and Z.W.\ contributed equally to the work, on the design and programming
of the FIREFLY code, and writing the initial draft of the paper.
H.-T.W.\ contributed to the ${\cal F}$-statistic-based method, and developed preliminary sampling codes for ringdown analysis in this work. 
J.Z.\ contributed to the Bayesian analysis and sampling implementation.
L.S.\ contributed to the design and methodology of FIREFLY, and writing of the paper, and provided supervision and coordination of the whole project.
All co-authors discussed the results and provided input to the data analysis and the content of the paper.



\clearpage

\begin{appendices}

\section{Zero-noise Injection}

In this Appendix, we present the results of \FF{} in zero-noise injections,
where the data are injected according to Eq.~(\ref{eq: ringdown QNM}) and
only involving the QNMs to be extracted. We show the test results in the
overtone analysis.  The source parameters $\bmt$ of the injected signals are
chosen to be the same as those in the main text, while the amplitudes and
phases of each QNM are set to reasonable values based on NR
analysis~\cite{Giesler:2019uxc}, as shown in Table \ref{tab:zero-noise
injection}.  Other configurations such as the priors and hyperparameters are
consistent with the overtone analysis in the main text. 

\begin{table}[h]
    \caption{Injected values of amplitudes $A_{lmn}$ and initial phases
    $\phi_{lmn}$ in zero-noise injection for different scenarios. Modes $N=1$,
    $N=2$, and $N=3$ correspond to the cases of injecting 1, 2, and 3 QNMs,
    respectively. The amplitudes and phases in the table denote the amplitude
    and initial phase of the signal at the beginning of the truncated starting
    times at $28\,{\rm M}$, $23\,{\rm M}$, and $18\,{\rm M}$ for the cases
    $N=1$, $N=2$, and $N=3$, respectively. The unit for $A_{\ell mn}$ is
    $10^{-20}$.}
    \label{tab:zero-noise injection}
    \begin{tabularx}{\textwidth}{XXXXXXX}
    \toprule
    Mode & $A_{220}$ & $\phi_{220}$ & $A_{221}$ & $\phi_{221}$ & $A_{222}$ & $\phi_{222}$\\ 
    \midrule
    $N=1$ & $0.1102$ & $5.4412$ & --      & --      & --  & --    \\
    $N=2$ & $0.1653$ & $2.7956$ & $0.0164$ & $5.2472$ & --  & --    \\
    $N=3$ & $0.2479$ & $0.1500$ & $0.0560$ & $2.6600$ & $0.0072$  & $1.5400$ \\
    \botrule
    \end{tabularx}
\end{table}

\begin{figure}[h]
    \centering
    \includegraphics[width=0.5\textwidth]{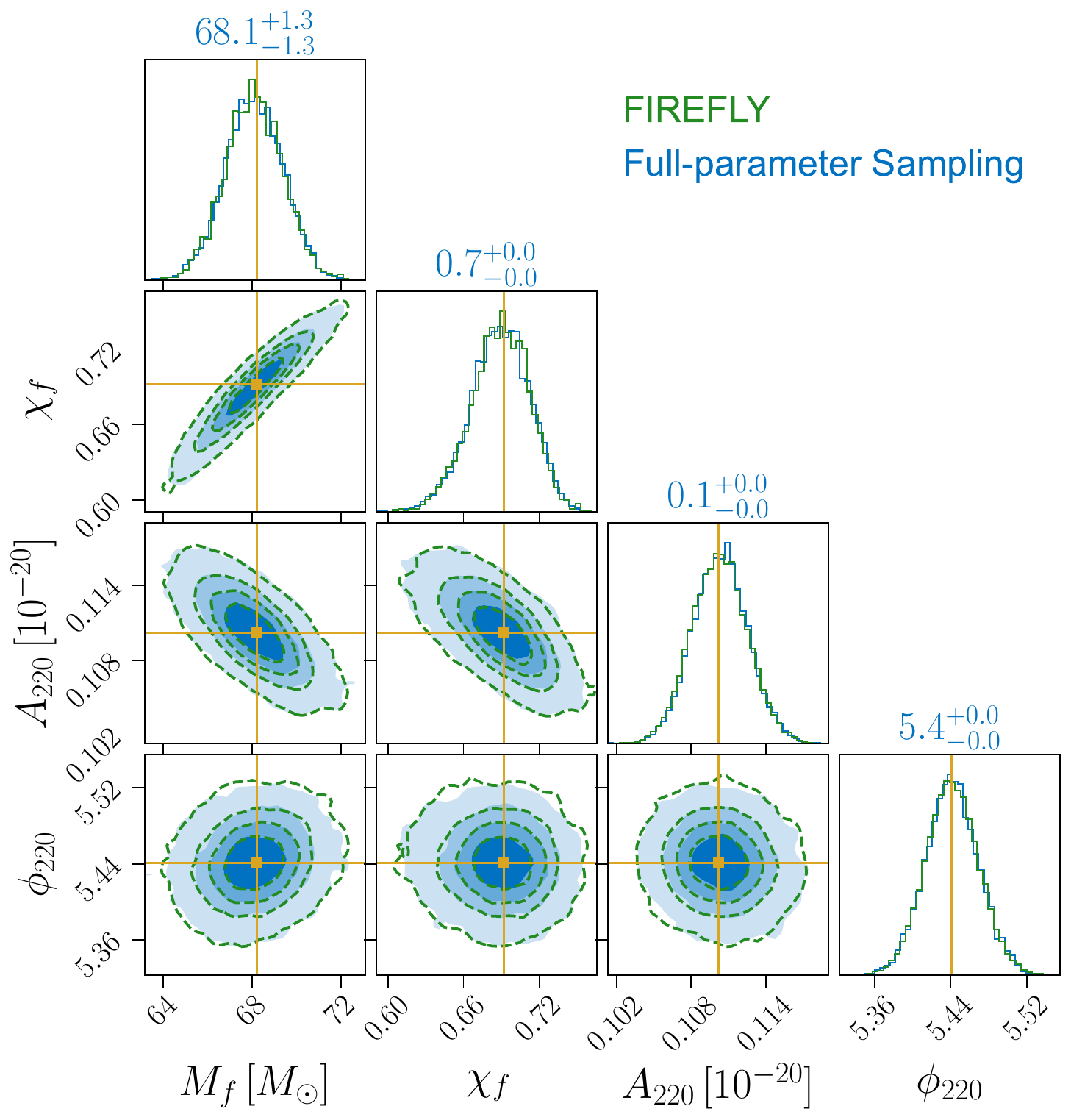}
    \caption{\textbf{Posteriors of the $N=1$ scenario in the zero-noise
    injection.} The contour levels and styles are the same as those in
    Fig.~\ref{fig2: posterior N=3}, while the data are injected with only the
    fundamental mode.}
\label{figA1: Posterior 221 ZN}
\end{figure}

\begin{figure}[h]
    \centering
    \includegraphics[width=0.7\textwidth]{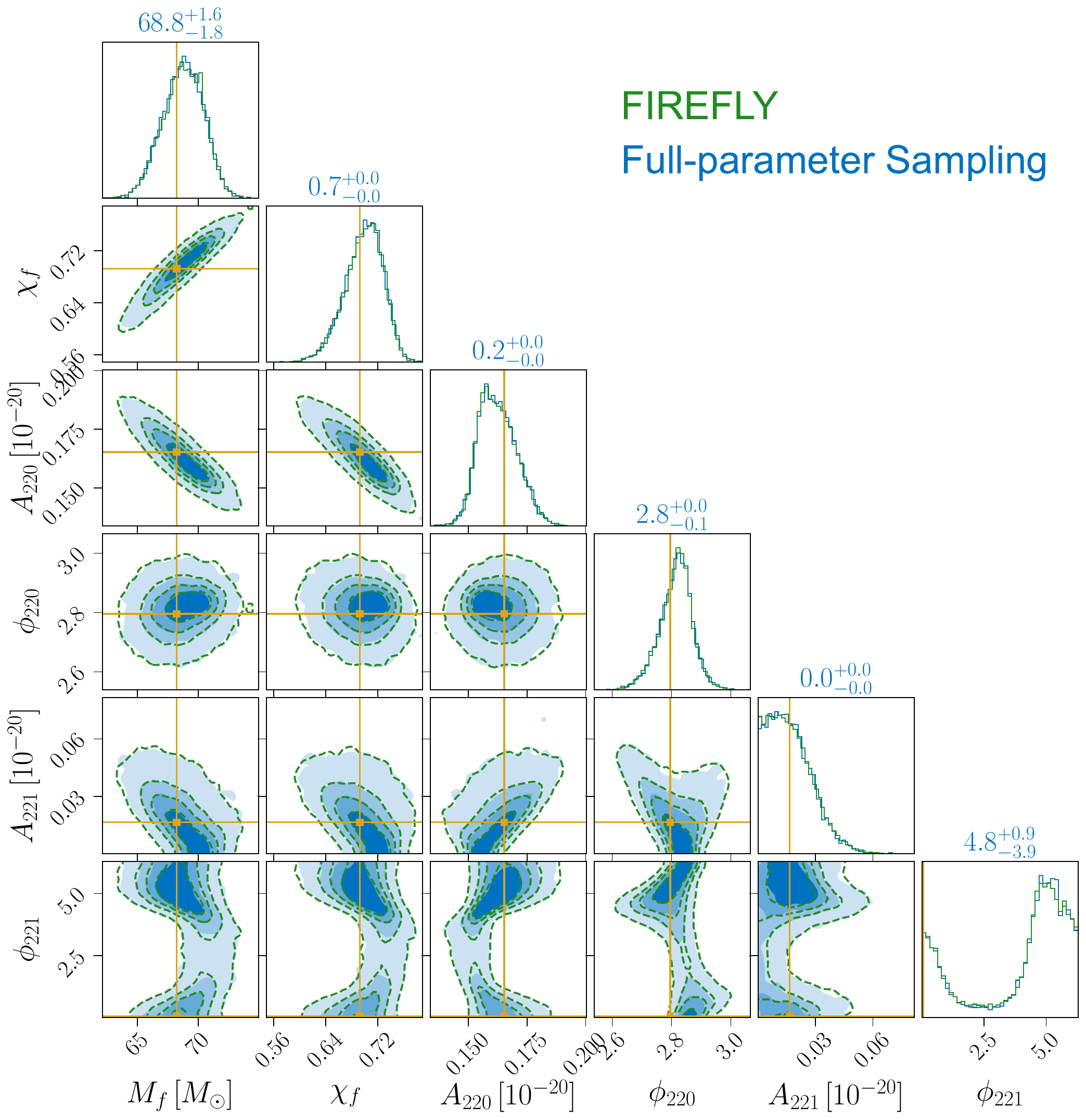}
    \caption{\textbf{Posteriors of the $N=2$ scenario in the zero-noise
    injection.} Same as Fig.~\ref{figA1: Posterior 221 ZN}, but for the $N=2$
    case with the fundamental and first overtone modes injected.}
\label{figA2: Posterior 222 ZN}
\end{figure}

\begin{figure}[h]
    \centering
    \includegraphics[width=1\textwidth]{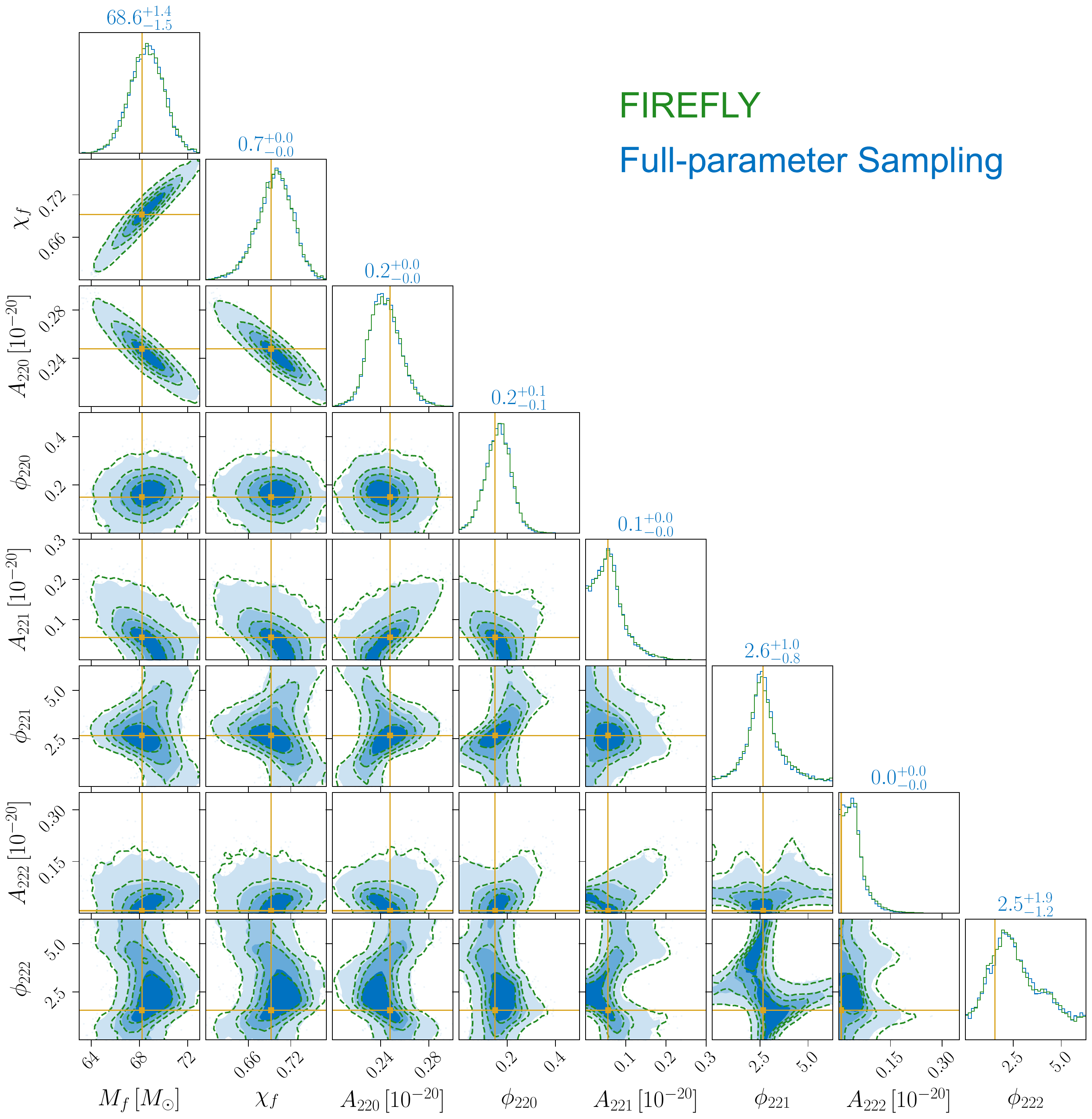}
    \caption{\textbf{Posteriors of the $N=3$ scenario in the zero-noise
    injection.} Same as Fig.~\ref{figA1: Posterior 221 ZN}, but for the $N=3$
    case with the fundamental, first overtone, and second overtone modes
    injected.}
\label{figA3: Posterior 223 ZN}
\end{figure}

\begin{figure}[h]
    \centering
    \includegraphics[width=1\textwidth]{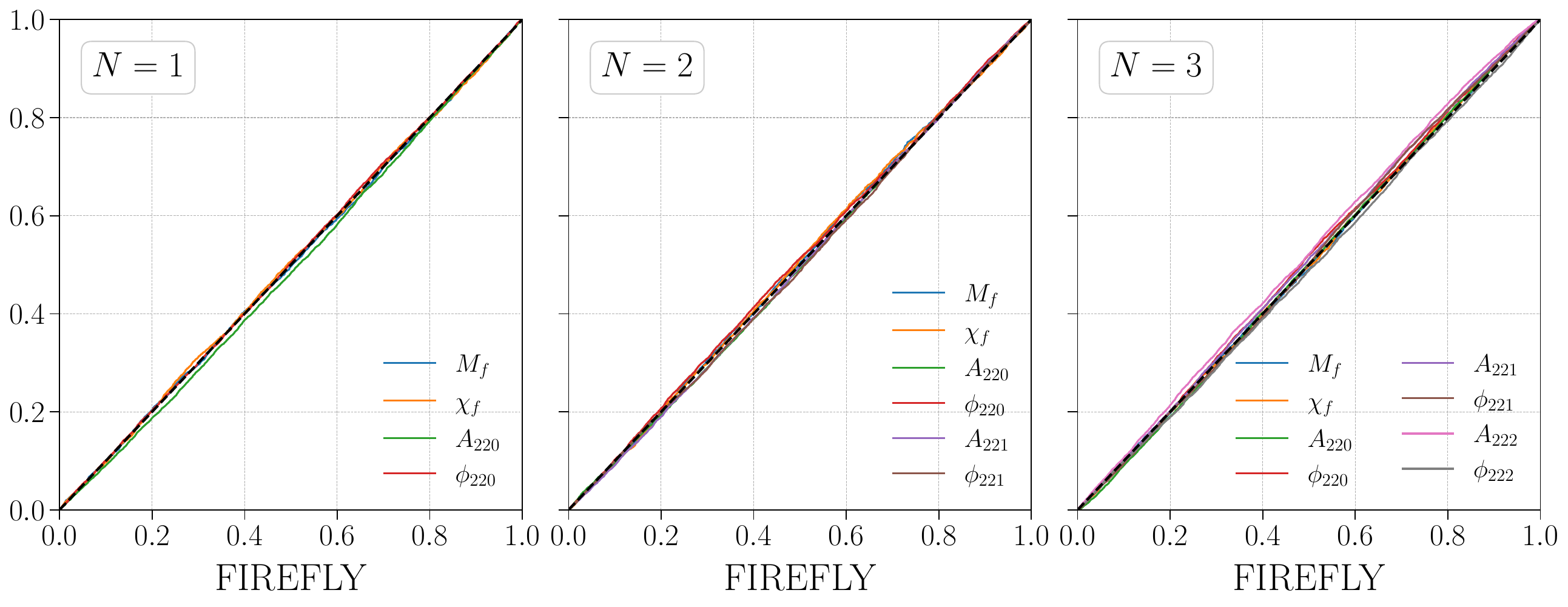}
    \caption{\textbf{P-P plots in the zero-noise injections.} Same as
    Fig.~\ref{fig3: P-P plot}, but for the zero-noise injections, where the
    data are injected with only the QNM signals under consideration.}
\label{figA4: P-P plot}
\end{figure}

\begin{figure}[h]
    \centering
    \includegraphics[width=1\textwidth]{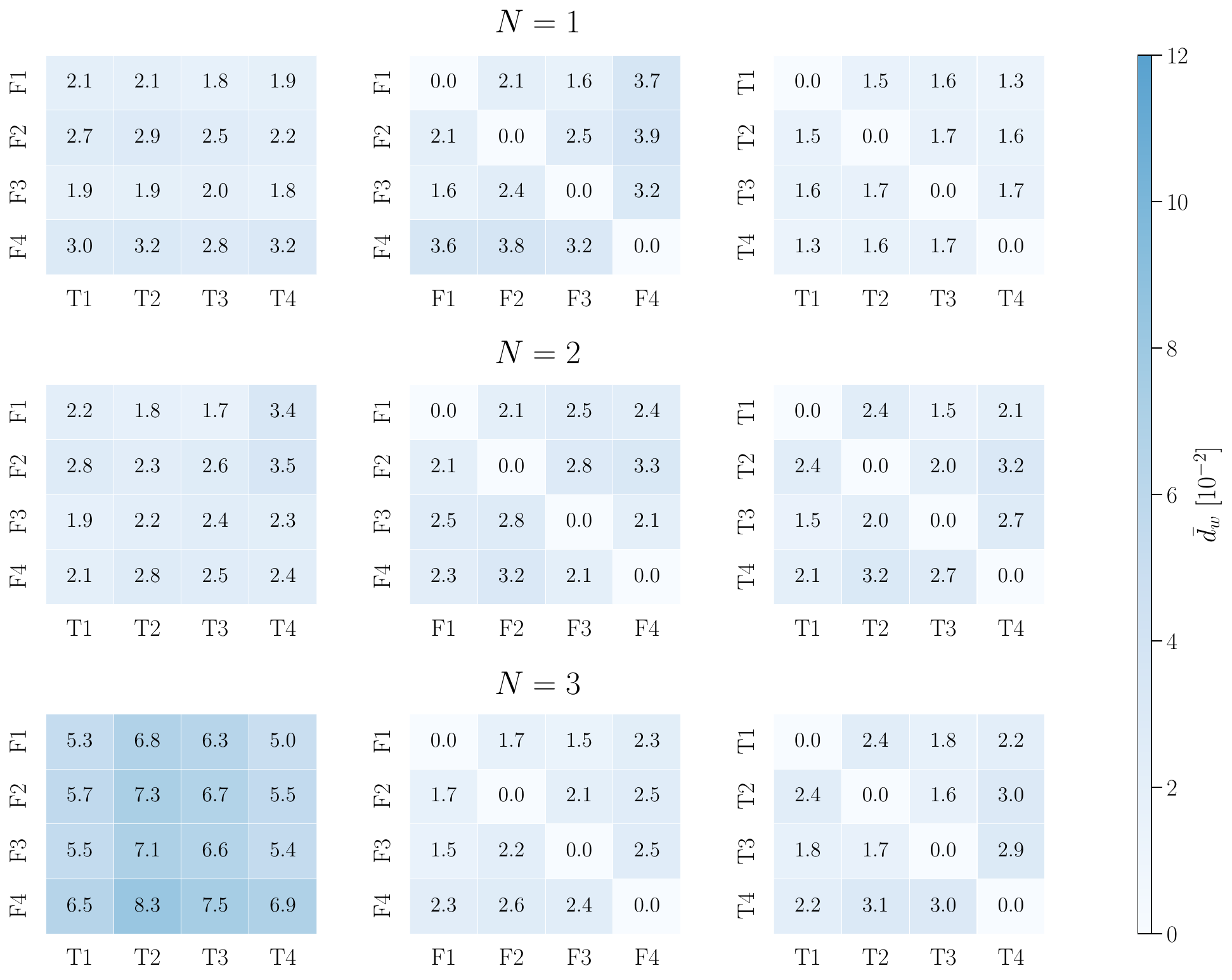}
    \caption{\textbf{Averaged and normalized one-dimensional Wasserstein
    distances for the posteriors of two methods in the zero-noise injections.}
    The three rows from top to bottom represent the results for $N=1$, $N=2$,
    and $N=3$ scenarios in overtone analysis. In each row, the three subplots
    from left to right show the comparisons between FIREFLY and the
    full-parameter sampling, the comparisons within FIREFLY, and the
    comparisons within the full-parameter sampling. In each subplot, the rows
    labeled from ``F1'' to ``F4'' represent four independent inferences using
    FIREFLY, while the columns labeled from ``T1'' to ``T4'' represent four
    independent inferences in the full-parameter sampling.  Each element in the
    matrix represents the averaged and normalized Wasserstein distance between
    the one-dimensional marginal posteriors.}
\label{figA5: WD ZN}
\end{figure}

The posteriors given by \FF{} and the full-parameter sampling are shown in
Figs.~\ref{figA1: Posterior 221 ZN}--\ref{figA3: Posterior 223 ZN},
corresponding to the cases of $N=1$, $N=2$, and $N=3$, respectively. While all results of the two
methods are highly consistent visually with each other and the injection
values, we further quantify the differences between the two methods using the
P-P plot and the Wasserstein distance. In Fig.~\ref{figA4: P-P plot}, we depict the P-P plot of
the one-dimensional marginal posteriors between the two methods in all three
scenarios.  The Wasserstein distance, also known as the Earth-Mover's
distance, measures the minimum cost of transforming one distribution into
another, providing a robust metric for comparing
distributions~\cite{MAL-073}. We compute the Wasserstein distance between
each pair of the one-dimensional marginal posteriors from the two methods,
and normalize it with the standard deviation of the one-dimensional posterior
distribution in the full-parameter sampling. The normalized Wasserstein
distances are further averaged over all parameters, providing a single,
average, normalized Wasserstein distance $\bar{d}_w$, shown in
Fig.~\ref{figA5: WD ZN}. For each model, the analysis is performed
independently four times for both \FF{} and the full-parameter sampling,
allowing for a comparison between the two methods as well as their internal
variability.  In the cases of $N=1$ and $N=2$, the differences between the
two methods are close to the internal fluctuations of each method, which
irrefutably demonstrates the accuracy of \FF{}. In the case of $N=3$, the
difference between the two methods is slightly larger than the internal
fluctuations, but remains at a relatively low level with $\bar{d}_{w}
\lesssim 0.1$, indicating that the average difference between the one-dimensional marginal distributions of the two methods is less than 0.1 standard deviation.

The evidences in different scenarios are also calculated by FIREFLY and the
full-parameter sampling. Same as in the main text, we conduct four independent
trials for each scenario, and show the results in Fig.~\ref{figA6: log evidence
ZN}. The dash-dotted lines represent the upper and lower bounds of the evidence
by taking the union of the ranges from the four independent trials. For the
full-parameter sampling, we further consider the uncertainties due to
insufficient sampling in the QNM parameters. We perform the following numerical
experiments to estimate it.  First, if $M_f$ and $\chi_f$ are fixed, then
${\cal L}(d|M_f, \chi_f,\bmB)$ has a Gaussian form in $\bmB$. In the auxiliary
inference, we analytically marginalize $\bmB$ and obtain the marginal
likelihood. In the full-parameter sampling, $\bmB$ is treated as the same as
$\bmt$ and sampled. We find that the nested sampling method tends to
overestimate the marginal likelihood in the numerically stochastic sampling of
$\bmB$. To quantify this bias, we fix $M_f$ and $\chi_f$ at their injected
values, and compute the Gaussian integration in $\bmB$ both analytically and
with the nested sampling method. The difference between the two results can be
regarded as the bias led by insufficient sampling in $\bmB$, shown in
Fig.~\ref{figA7: Gaussian integral ZN}. For the higher-dimensional cases of
$N=2$ and $N=3$, the nested sampling tends to produce overestimated values,
which partially explains the small evidence discrepancy between the two methods
discussed in the main text. Without further investigation, in this work we
simply treat the bias as an additional uncertainty in the full-parameter
sampling, and combine it with the evidence uncertainty given by the
nested-sampling algorithm. This finally leads to a more conservative estimation
of the evidence.

\begin{figure}[h]
    \centering
    \includegraphics[width=0.5\textwidth]{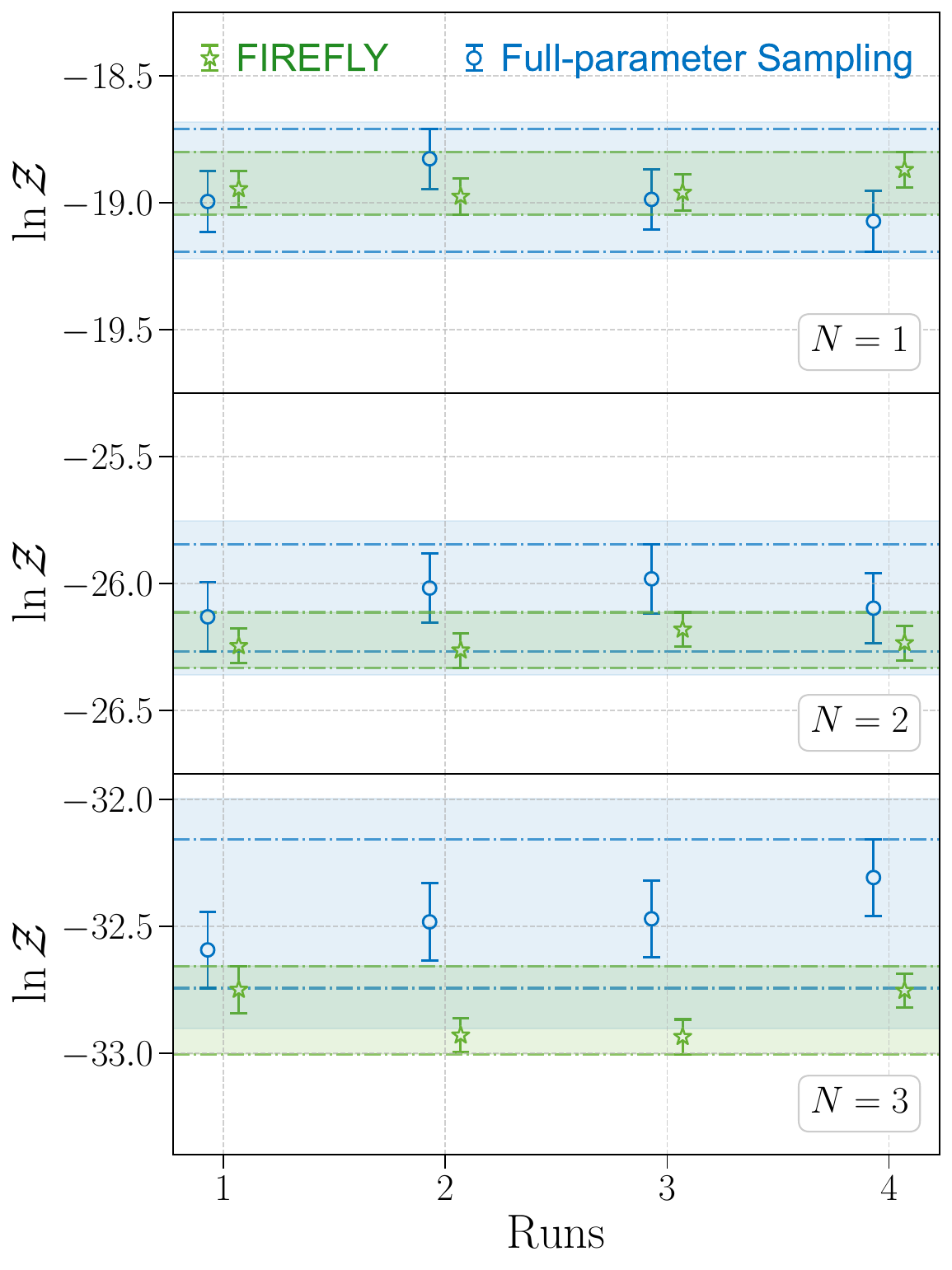}
    \caption{\textbf{Evidences calculated by FIREFLY and the full-parameter
    sampling in the zero-noise injections.} Same as Fig.~\ref{fig4: log
    evidence}, but for the zero-noise injections.}
\label{figA6: log evidence ZN}
\end{figure}
\begin{figure}[h]
    \centering
    \includegraphics[width=0.5\textwidth]{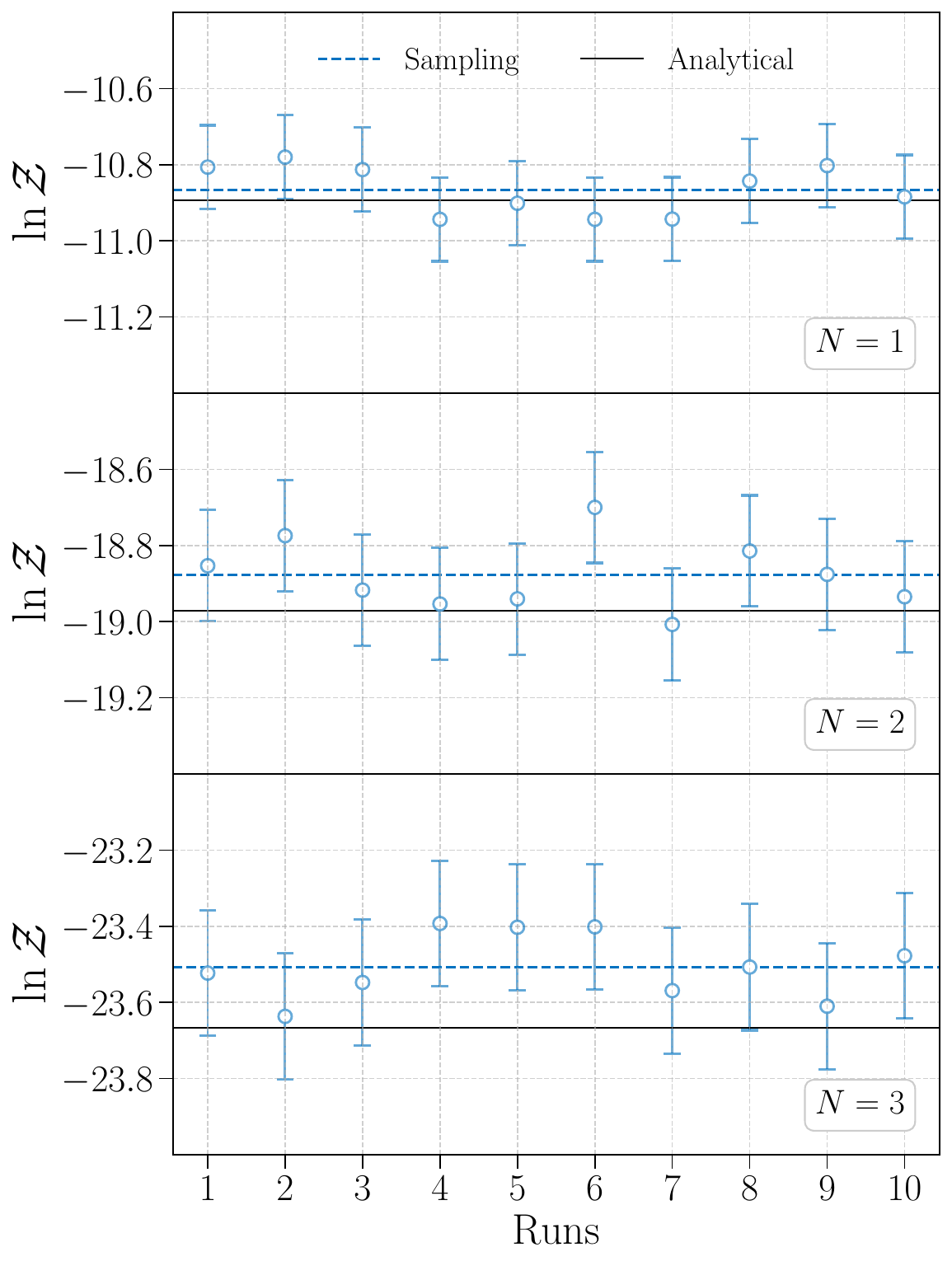}
    \caption{\textbf{Integrals of Gaussian functions over QNM parameters.} The
    subplots represent the integral over the QNM parameters $\bmB$, where $M_f$
    and $\chi_f$ are fixed at the injected values. From top to bottom, the
    three subplots display the results in the cases of $N=1$, $N=2$, and $N=3$,
    with 2, 4, and 6 QNM parameters, respectively. For each scenario, we run
    the nested sampling to calculate the integral in $\bmB$ for 10 times. The
    results are shown as dots, with their mean as black horizontal lines. The
    gray horizontal lines show the analytical values for comparison.}
\label{figA7: Gaussian integral ZN}
\end{figure}






\end{appendices}

\clearpage
\bibliography{sn-bibliography}
\end{document}